\documentclass[11pt]{article}

\usepackage{alltt}
\usepackage{amsmath}
\usepackage{amsthm}
\usepackage{booktabs}
\usepackage{graphicx}
\usepackage{listings}
\usepackage{multirow}
\usepackage{microtype}
\usepackage{makecell}
\usepackage{url}
\usepackage{colortbl}
\usepackage{subcaption}

\usepackage[preprint]{automl}
\usepackage{cleveref}
\crefname{lstlisting}{listing}{listings}
\Crefname{lstlisting}{Listing}{Listings}

\usepackage[square,numbers]{natbib}
\title{Resolving Java Code Repository Issues with iSWE Agent}

\author[1]{\nameemail{Jatin Ganhotra}{jatinganhotra@us.ibm.com}}
\author[1]{\nameemail{Sami Serhan}{Sami.Serhan@ibm.com}}
\author[1]{\nameemail{Antonio Abu Nassar}{Antonio.Abu.Nassar@ibm.com}}
\author[1]{\\\nameemail{Avraham Shinnar}{shinnar@us.ibm.com}}
\author[1]{\nameemail{Ziv Nevo}{nevo@il.ibm.com}}
\author[1]{\nameemail{Martin Hirzel}{hirzel@us.ibm.com}}

\affil[1]{IBM}

\hypersetup{%
  pdfauthor={}, % will be reset to "Anonymous" unless the "final" package option is given
  pdftitle={},
  pdfsubject={},
  pdfkeywords={}
}

\definecolor{keyword}{HTML}{37AC4A}
\definecolor{operator}{HTML}{A51DFF}
\definecolor{string}{HTML}{0070C0}

\lstdefinelanguage{issue}{
  morecomment=[s]{```}{```},
  commentstyle=\color{string},
}
\lstdefinelanguage{localization}{
  morestring=[b]",
  stringstyle=\color{string}\rm,
  literate=
    {[}{{\bf\texttt{\color{keyword}{[}}}}{1}
    {]}{{\bf\texttt{\color{keyword}{]}}}}{1}
    {\{}{{\bf\texttt{\color{keyword}{\char '173}}}}{1}
    {\}}{{\bf\texttt{\color{keyword}{\char '175}}}}{1},
}
\lstdefinelanguage{edit}{
  morecomment=[s]{<<<<<<<}{REPLACE},
  commentstyle=\color{string},
}

\begin{document}

\maketitle

\begin{abstract}
Resolving issues on code repositories is an important part of
software engineering.
Various recent systems automatically resolve issues using large
language models and agents, often with impressive performance.
Unfortunately, most of these models and agents focus primarily on
Python, and their performance on other programming languages is lower.
In particular, a lot of enterprise software is written in Java,
yet automated issue resolution for Java is underexplored.
This paper introduces iSWE Agent, an automated issue resolver
with an emphasis on Java.
It consists of two sub-agents, one for localization and the other
for editing.
Both have access to novel tools based on rule-based Java static
analysis and transformation.
Using this approach, iSWE achieves state-of-the-art issue
resolution rates across the Java splits of both Multi-SWE-bench and
SWE-PolyBench.
More generally, we hope that by combining the best of rule-based and
model-based techniques, this paper contributes towards improving
enterprise software development.
\end{abstract}

\section{Introduction}\label{sec:introduction}

One way to make software developers more productive is by giving them
automated support for resolving issues.
An \emph{issue} on a code repository is a natural-language bug report
or feature request, and resolving an issue involves making changes to
the code that fix the bug or implement the feature.
The SWE-bench benchmark~\cite{jimenez_et_al_2024} measures the ability
of automated systems to resolve Python issues and has inspired many,
mostly agent-based, solutions using large language models~(LLMs).
However, while Python-based issue resolution leaderboards are highly
active and show signs of saturating, issue resolution leaderboards for
other programming languages are less active and show lower performance.
At the same time, other programming languages are of great practical
importance: for instance, much enterprise software is written in Java.
As a language, Java differs substantially from Python in terms of its
syntax, type system, and emphasis on object-orientation.
This raises the question of whether Java issue resolution would
benefit from Java-specific knowledge~--- a question that is, to our
knowledge, unanswered.
This paper sets out to tackle that gap.

Shortly after the release of SWE-bench, SWE-Agent demonstrated early
gains thanks to carefully crafted tools for performing code-related
actions~\cite{yang_et_al_2024}.
In particular, its edit tool had a built-in Python linter for catching
coding mistakes, suggesting that language-specific knowledge can be
helpful.
At the other extreme, the CodeAct paper~\cite{wang_et_al_2024} argued
against sophisticated tools, instead expressing all actions purely as
code.
And recent frontier models have improved SWE-bench success rates for
Python even when the action space only allows code in the bash shell
language.
Unfortunately, such arbitrary code actions risk unsafe side-effects,
and avoiding those side-effects via sandboxing incurs overheads in
time, memory, disk, and solution complexity.
Arbitrary code actions also tend to require more iterations of the
agentic loop, increasing LLM inferencing cost compared to tools
tailored towards issue resolution for a given language.
Therefore, this paper explores the premise that we can improve an
agent for Java issue resolution by giving it advanced Java-specific
tools.

This paper introduces iSWE, a new LLM-based agent for resolving issues
in Java code repositories.
(While iSWE can also handle other languages besides Java, it does so
in a more generic manner.)
Compared to other issue-resolution agents, iSWE puts an emphasis on
stronger tools, meaning tools that are:
(i)~more effective by leveraging Java-specific knowledge;
(ii)~less LLM-dependent and able to use fewer LLM turns by providing a
simple high-level interface;
(iii)~safer by avoiding spurious side-effects;
and
(iv)~lighter-weight by only launching a sandbox for rare actions where
side-effects are unavoidable.
Internally, iSWE comprises two ReAct~\cite{yao_et_al_2023} sub-agents,
one for localization and the second for editing.
Decomposing issue resolution this way makes each sub-task easier, and
has the added benefit that even solving just localization on its own
delivers incremental value to developers.
The sub-agents, in turn, build upon two open-source technologies: the
declarative prompting language PDL~\cite{vaziri_et_al_2024} and the
program analysis framework CLDK~\cite{krishna_et_al_2025}.

We evaluate iSWE on the Java subsets of two issue-resolution
benchmarks, Multi-SWE-bench~\cite{zan_et_al_2025} and
SWE-PolyBench~\cite{rashid_et_al_2025}.
We selected these two because both have public leaderboards, and
because taken together, they have enough Java instances to draw
meaningful conclusions~(128+165=293).
The evaluation demonstrates iSWE's performance and sheds light on some
idiosyncracies of the benchmarks that are of interest to the
issue-resolution research community.
In terms of success rate, iSWE ranks at or near the top for Java
on both leaderboards.
And in terms of cost, iSWE spends between 2$\times$ and 3$\times$
fewer dollars on model-inferencing APIs than other leading agents
using the same LLM.
The results section further explores other metrics (e.g.,~localization
precision and recall) and breakdowns (e.g.,~by issue complexity).
Overall, this paper makes the following contributions:

\begin{enumerate}
  \item It introduces iSWE, the first issue-resolution agent
    specialized for Java.
  \item It explores the premise that language-aware tools can yield
    stronger issue-resolution agents.
  \item It reports on extensive experiments with two thus-far less
    explored Java SWE benchmarks.
\end{enumerate}

We hope this paper contributes towards taking some of the research
progress driven by Python-centered benchmarks and extending its
benefits into enterprise languages such as Java.

\section{Background}\label{sec:background}

The \emph{problem statement} for issue resolution is as follows:
\begin{quote}
Given an issue description~$d_\textrm{issue}$ and the current code in
a repository~$c_\textrm{old}$, generate a new modified
version~$c_\textrm{new}$ of the code that resolves the issue.
\end{quote}

\begin{lstlisting}[float, language=issue, caption={Example issue description $d_\textrm{issue}$, based on SWE-PolyBench instance \textit{apache\_\_rocketmq-516}.}, label=lst:issuedesc]
org.apache.rocketmq.common.stats.StatsItemSet#getAndCreateStatsItem has multi-thread problem.
Because we can not ensure the atomicity of
```java
StatsItem statsItem = this.statsItemTable.get(statsKey);
StatsItem prev = this.statsItemTable.put(statsKey, statsItem);
```
Here is the test case. The result is not always the correct one 20000.
```java
for(int i =0; i < 10000; i++) {
  executor.submit(new Runnable() {
    @Override
    public void run() {
      brokerStatsManager.incTopicPutNums("topicTest", 2, 1);
    }
  });
}
Thread.sleep(5000);
System.out.println(brokerStatsManager.getStatsItem(TOPIC_PUT_NUMS,"topicTest").getValue());
```
\end{lstlisting}

\Cref{lst:issuedesc} shows an example issue description~$d_\textrm{issue}$.
This particular issue description indicates a code location to fix,
though it will turn out that properly resolving the issue also
requires changing code in a second location not mentioned in the issue.
Most issue descriptions do not clearly indicate code locations.
In real-world practice, after a developer writes new modified code~$c_\textrm{new}$, they usually do some testing to convince themselves that it addresses
the issue, then submit a pull request~(PR) to update the repository.
Issue description often include an informal acceptance test fragment,
as in the example in \Cref{lst:issuedesc}.
Sometimes, developers even include a more polished version of such a
test in the same PR along with the issue-resolving code~$c_\textrm{new}$.
Benchmarks like SWE-bench~\cite{jimenez_et_al_2024} emulate this
practice by evaluating each candidate~$c_\textrm{new}$ on a hidden set
of test cases~$t_\textrm{gold}$.
This yields an execution-based evaluation, accounting for the fact
that what matters is whether generated code resolves the issue, not
whether it resembles ground-truth code.

What makes SWE-bench and similar benchmarks realistic is that they are
based on (often popular) open-source code repositories:
they mine issue-resolving PRs to get an issue~($d_\textrm{issue}$)
along with the code before~($c_\textrm{old}$) and
after~($c_\textrm{new}$) the PR.
For example, \Cref{lst:issuedesc} is mined from the
\textit{apache/rocketmq} repository.
Then, they filter for presence of tests~($t_\textrm{gold}$) that are
fail-to-pass~(F2P), i.e., they fail on~$c_\textrm{old}$, reproducing the issue, and pass on~$c_\textrm{new}$, confirming its resolution.
For completeness, in addition to F2P tests, the benchmark also
runs a select few pass-to-pass~(P2P) tests to check whether the
patch broke any previous behavior.
SWE-bench Verified~\cite{chowdhury_et_al_2024} is a subset of 500
SWE-bench instances filtered further using human annotations to ensure issue
descriptions are not underspecified and tests are not overly specific.

While SWE-bench and SWE-bench Verified focus on Python, there are at
least three benchmarks that address the same problem statement but for
other programming languages including Java.
\emph{Multi-SWE-bench}~\cite{zan_et_al_2025} comprises 1,632~instances
from 7~languages, including 128~instances from~Java.
All its instances are pre-filtered using the same criteria as for
SWE-bench Verified, and further hand-labeled for difficulty levels.
\emph{SWE-PolyBench}~\cite{rashid_et_al_2025} comprises
2,110~instances from 4~languages, including 165~instances from~Java.
It comes with complexity labels based on how many functions and
classes the ground-truth PR changed.
SWE-PolyBench Verified is a hand-filtered subset, including
69~instances from Java.
\emph{SWE-bench Multilingual}~\cite{yang_et_al_2025} comprises
300~instances from 9~languages, including 43~instances from~Java.

This paper adopts both Multi-SWE-Bench and SWE-PolyBench for
evaluation to increase the number and diversity of Java instances.
Doing so also enables comparing iSWE against other submissions to the
corresponding public leaderboards, since several systems have
submissions on one or the other but not both.
Furthermore, it reduces the risk of inadvertently tailoring the agent
scaffold too much to one particular benchmark, thus facilitating
practical use for other repositories beyond the benchmarks.
This paper does not adopt SWE-bench Multilingual, since it only has
a small number of additional Java instances.

Issue resolution tasks for Java differ from those from Python
SWE-bench in several ways.
Python agents tend to struggle with multi-file edits~\cite{ganhotra2025multifile},
and  golden code edits~$c_\textrm{new}$ for Java issues tend to affect
more files and involve more diff-hunks than those for Python issues.
This may be caused by Java's object-oriented nature, which encourages
code where functionality is spread across more files.
Also, Java is a compiled language with a strong static type systems.
Since many code mistakes are caught by the Java compiler, the popular
Java linters do not need to check for those same mistakes.
As a consequence, unlike for Python, an automatic issue resolution
system for Java cannot rely on a linter alone to catch mistakes.
Furthermore, the Java compiler requires the correct versions of
libraries it depends on in the build environment, and most projects
invoke the Java compiler from a build tool such as Gradle or Maven.
Taken together, this makes it harder for an automatic issue resolution
system to get feedback from static correctness checks.

Another way in which issue resolution for Java differs from Python is
that the Python leaderboards are more popular and are starting to show
signs of saturation~\cite{ganhotra2025saturation}.
Some voices~\cite{badertdinov_et_al_2025,deng_et_al_2025,liang_garg_moghaddam_2025,openai_verified_no_longer_2026}
even raise contamination concerns with Python SWE-bench, more likely due
to extended pre-training than to fine-tuning.
That said, leading frontier-model vendors are indeed fine-tuning their LLMs
for Python issue resolution~\cite{baker_et_al_2025,macdiarmid_et_al_2025}.
While this fine-tuning hopefully uses data disjoint from the benchmark
test set, it is nevertheless likely to improve the model more for
Python tasks than for other languages such as Java.
On the other hand, Java issue resolution has received relatively less
attention so far, making it less saturated and less overfitted.

\section{Approach}\label{sec:approach}

\begin{figure}
  \centerline{\includegraphics[width=0.9\textwidth]{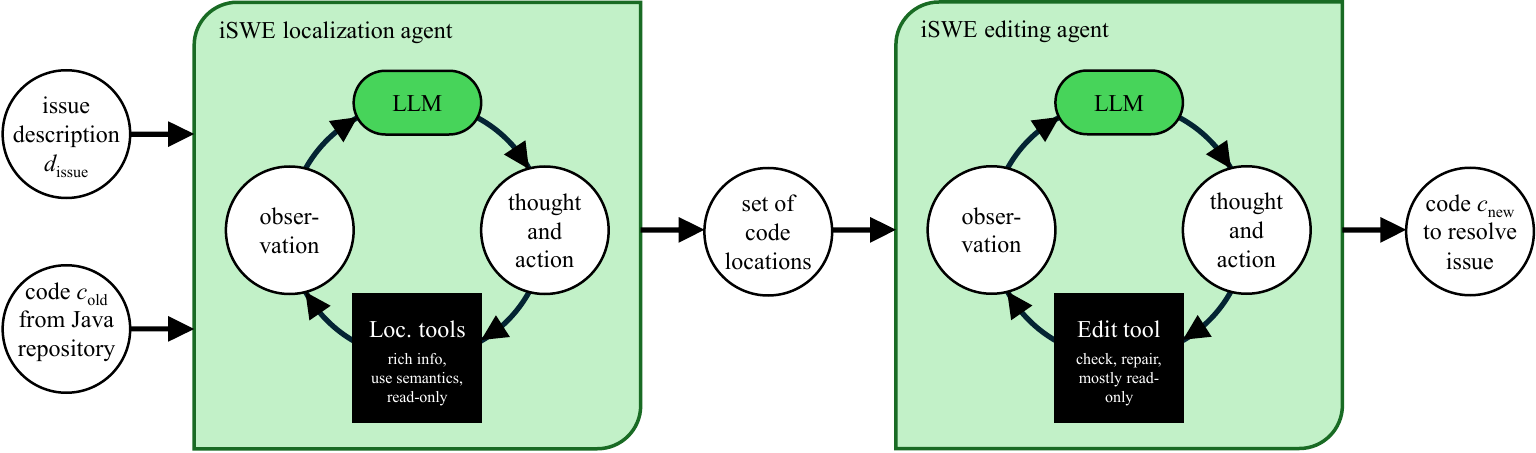}}
  \caption{\label{fig:overview}Overview of iSWE Agent}
\end{figure}

This section describes our approach for tackling the problem statement
from Section~\ref{sec:background}, namely, given an issue issue
description~$d_\textrm{issue}$, mapping from old code~$c_\textrm{old}$
to new code~$c_\textrm{new}$.
\Cref{fig:overview} shows our two-agent pipeline for solving
this problem.
The pipeline first feeds the input
\mbox{$\langle d_\textrm{issue},c_\textrm{old}\rangle$}
into the localization agent.
The localization agent is a ReAct~\cite{yao_et_al_2023} agent with
read-only tools, and it returns an edit recommendation with a set
of code locations in~$c_\textrm{old}$
that should be edited to resolve the issue.
Next, the pipeline feeds this recommendation and set of code locations
into the editing agent.
The editing agent is also a ReAct agent with a mostly read-only tool,
and it returns the modified code~$c_\textrm{new}$ that, if correct,
resolves the issue.

There are several reasons why we opted for this flow of two ReAct agents.
One advantage is that it gives us the flexibility to run either the entire
flow end-to-end or each sub-agent separately.
For instance, sometimes users want to pause after the first step,
to inspect the intermediate result.
Conversely, sometimes the first step can be skipped, if the localization
is available from a different source such as a developer or a static
code scanning tool.
The intermediate result itself is typed, structured, and sanitized.
This simplifies our engineering effort, since each sub-agent is modular
with well-defined input and output expectations.
It also speeds up experiments, since we can run pieces in isolation.
Finally, each of the sub-tasks is relatively simpler than the
end-to-end task, bringing it more within reach of moderate-sized LLMs.

Both agents are LLM-agnostic: they avoid hardwired LLMs or LLM-specific
prompts, and instead, the LLM to be called is a configuration variable.
As of now, iSWE uses \emph{inline reasoning}, in the sense that the LLM may
generate a chain-of-thought~\cite{wei_et_al_2022} as part of its main content, as opposed to
generating a separate API response field such as \textit{reasoning\_content}.
Similarly, iSWE uses \emph{inline tool calling}, in the sense that the LLM
may generate a tool call~\cite{schick_et_al_2023} as part of its main content, which iSWE then
parses on the client-side, as opposed to using a native tool-calling API.
We found these inline approaches to be more robust across different LLMs.

We implemented both agents using Python and
PDL~\cite{vaziri_et_al_2024,spiess_et_al_2025}.
PDL is a declarative YAML-based language for writing prompt templates
and the core LLM-flow control logic directly surrounding the prompts.
It takes care of context accumulation and data flow between
components including type-checking.
By using PDL, iSWE keeps its prompts easy-to-read, compared to the
common practice of scattering their sub-strings across Python code.
At the same time, PDL is flexible enough to work for a variety of
different flows, and thus, we did not have to work around design
decisions of an existing agentic framework such as
AutoGen~\cite{wu_et_al_2023} or OpenHands~\cite{wang_et_al_2025}.
In particular, one core decision we wanted control over was
containerization.
Most parts of iSWE run in user-space without any container, including
the top-level flow, the ReAct loop of each of the sub-agents, and even
the functions being called for tools.
This makes things simpler and faster.
The only part that runs in a container is the part of the editing
agent that builds the Java project for the purpose of error
checking~(described below).
The following two subsections describe each of iSWE's two sub-agents,
localization and editing, in detail.

\subsection{iSWE Localization Agent}

\begin{lstlisting}[float, language=localization, caption={Example sanitized localization for the issue description from \Cref{lst:issuedesc}.}, label=lst:localization]
{thought: "Perfect! Now I have all the information needed to identify the bug and propose fixes. The issue is a race condition in the getAndCreateStatsItem method in both StatsItemSet and MomentStatsItemSet classes.",
 localization: {
   "main/java/org/apache/rocketmq/common/stats/MomentStatsItemSet.java": [
     {changes: [{kind: "modify", range: {start: 76, end: 90}}],
      proposed_edit: "Replace the non-atomic get-check-put pattern with putIfAbsent() to ensure atomicity.",
      scope: [
         {name: "MomentStatsItemSet",
          kind: {language: "java", name: "class" },
          span: {start: 28, end: 92}},
         {name: "getAndCreateStatsItem",
          kind: { language: "java", name: "method" },
          span: {start: 76, end: 91}}]
     }
   ],
   "main/java/org/apache/rocketmq/common/stats/StatsItemSet.java": [
     {changes: [{kind: "modify", range: {start: 160, end: 173}}],
      proposed_edit: "Replace the non-atomic get-check-put pattern with putIfAbsent() to ensure atomicity.",
       scope: [
         {name: "StatsItemSet",
          kind: {language: "java", name: "class" },
          span: {start: 28, end: 203}},
         {name: "getAndCreateStatsItem",
          kind: { language: "java", name: "method" },
          span: {start: 160, end: 174}}]
   }
  ]
}
\end{lstlisting}

As shown in \Cref{fig:overview}, the input to the localization agent
consists of the issue description~$d_\textrm{issue}$ and the old
code~$c_\textrm{old}$.
The initial prompt context before the first agent iteration
includes~$d_\textrm{issue}$ but not~$c_\textrm{old}$, which can be
very large; instead, the agent can inspect~$c_\textrm{old}$ via tools.
To this end, the initial prompt context also contains instructions and
tool descriptions.
Each loop iteration starts with an LLM call to produce a
thought~(inline reasoning) and an action~(tool call).
The agent parses the tool call, invokes the tool, and obtains the
output from the tool as an observation.
PDL~\cite{vaziri_et_al_2024} implicitly appends the
\mbox{$\langle$thought,action,observation$\rangle$}
triple from the current loop iteration to the prompt context, which
becomes the LLM input in the next iteration.
The loop terminates when the LLM generates not a tool call but a valid
JSON with a set of code locations.

\Cref{lst:localization} shows an example of a sanitized JSON output
from the localization agent, containing a set of code locations.
In this case, there are two locations across two different files.
Only one of them was mentioned in the corresponding issue
description~$d_\textrm{issue}$~(\Cref{lst:issuedesc}), demonstrating
that the localization agent found additional useful information.
Besides the locations, the JSON in \Cref{lst:localization} also
includes the final thought from the localization agent, which might be
useful for downstream consumption in the editing agent.
In our implementation, we only expect the LLM to generate a simplified
subset of the information in \Cref{lst:localization}, and then fill in
the remaining details using rule-based program analysis.
The LLM output JSON is less nested than the final sanitized JSON.
For example, line number ranges are given as a string such
as \lstinline[language=localization]{"76-90"}, and scopes do not
provide spans, or can even be elided altogether.
Then, the rule-based sanitizer adds any incomplete information and
reconciles contradictions, if any.
The initial LLM context contains few-shot examples for the LLM output
JSON format to teach the localization agent what it needs to produce.

By default, the localization agent is configured to use the following
seven tools:
\textit{get\_file\_info}, 
\textit{get\_class\_info}, 
\textit{get\_method\_info}, 
\textit{get\_symbol\_info}, 
\textit{get\_inheritance\_hierarchy}, 
\textit{get\_function\_callers}, and
\textit{get\_call\_chain}, 
If we conceptualize the code~$c_\textrm{old}$ as a graph, then the
first four tools get information on nodes representing files, classes,
methods, or other symbols, and the remaining three tools get
information on edges representing class inheritance and function calling.
We implemented these tools using two rule-based static analysis libraries,
CLDK~\cite{krishna_et_al_2025} and Tree-Sitter~\cite{clem_thomson_2022}.
All tools generate textual outputs carefully crafted to provide rich
yet concise information to the LLM.
Also, all of the above-listed tools are read-only, in the sense that
they only passively obtain information without making any changes
to~$c_\textrm{old}$ or having other side-effects on the system.
We used ACE~\cite{agarwal_et_al_2025} to hone the tool descriptions
to make it easier for LLMs to call them.

Above, we describe the localization agent in its default configuration.
Our implementation also permits various aspects to be configured
differently.
For instance, we can provide different tools, such as those from
SWE-Agent~\cite{yang_et_al_2024} or even a bash tool.
We can change the number of few-shot samples for the JSON format.
And we can configure the output of tools, for instance, whether to
include line numbers, or whether to show relative paths.

\subsection{iSWE Editing Agent}

\begin{lstlisting}[float, language=edit, label=lst:edit, caption={Example edit for the issue from \Cref{lst:issuedesc} and the localization from \Cref{lst:localization}.}]
The solution is to use `putIfAbsent()` which is an atomic operation. If `putIfAbsent()` returns non-null, it means another thread already inserted a value, so return that existing value
```java
### main/java/org/apache/rocketmq/common/stats/MomentStatsItemSet.java
<<<<<<< SEARCH
  MomentStatsItem statsItem = statsItemTable.get(statsKey);
  if (null == statsItem) {
    statsItem = new MomentStatsItem(statsName, statsKey, schedExSvs, log);
    MomentStatsItem prev = statsItemTable.put(statsKey, statsItem);
=======
  MomentStatsItem statsItem = statsItemTable.get(statsKey);
  if (null == statsItem) {
    statsItem = new MomentStatsItem(statsName, statsKey, schedExSvs, log);
    MomentStatsItem prev = statsItemTable.putIfAbsent(statsKey, statsItem);
    if (null != prev) statsItem = prev;
>>>>>>> REPLACE
### main/java/org/apache/rocketmq/common/stats/StatsItemSet.java
<<<<<<< SEARCH
  StatsItem statsItem = statsItemTable.get(statsKey);
  if (null == statsItem) {
    statsItem = new StatsItem(statsName, statsKey, schedExSvs, log);
    StatsItem prev = statsItemTable.put(statsKey, statsItem);
=======
  StatsItem statsItem = statsItemTable.get(statsKey);
  if (null == statsItem) {
    statsItem = new StatsItem(statsName, statsKey, schedExSvs, log);
    StatsItem prev = statsItemTable.putIfAbsent(statsKey, statsItem);
    if (null != prev) statsItem = prev;
>>>>>>> REPLACE
```
\end{lstlisting}

The input to the editing agent consists of the set of locations
returned by the localization agent.
Based on this set of locations, the agent extracts and formats
relevant excerpts from~$c_\textrm{old}$, and places them into the
initial prompt context.
For example, given the localization in \Cref{lst:localization}, it
would show excerpts from two Java files, using the line ranges of
the \lstinline[language=localization]{scope} fields to show only the
methods where edits should take place.
Since code without line numbers is more in-distribution for most
LLMs than code with line numbers, the agent does not explicitly
mention line numbers in the prompt.
Instead, it injects \emph{edit marker} comments to pinpoint exact
locations to edit.
Besides excerpts from~$c_\textrm{old}$, the initial prompt before the
first agent iteration also includes the issue description~$d_\textrm{issue}$,
the \lstinline[language=localization]{thought}
and \lstinline[language=localization]{proposed_edit} fields from the
localization, plus instructions and a description of the edit tool.

Similarly to the localization agent, each loop iteration of the
editing agent also starts with an initial LLM call to produce a
thought and an action.
The action is an edit tool call, which comprises one or more
search-replace blocks, across one or more files.
The agent parses the tool call from the LLM output and tries to apply
it to an in-memory copy of the relevant portion of $c_\textrm{old}$ to
obtain a candidate~$c_\textrm{new}$.
Since the original version of $c_\textrm{old}$ on disk remains
unchanged, this step is trivially free of side effects and requires no
container sand-box.
If the editing tool detects any mistakes in the tool call, it renders
them as an observation.
The previous context plus the
current \mbox{$\langle$thought,action,observation$\rangle$} triple
become the input to the LLM call in the next loop iteration.
The loop terminates when the editing tool cannot find any mistakes in
the tool call, at which point the agent returns the
well-formed~$c_\textrm{new}$, converted to unidiff format.

\Cref{lst:edit} shows an example of an edit tool call, based on the
same running example used in earlier sections.
It takes the form of a search-replace in \emph{merge-conflict format}.
This format has been used for many years by version management
software~\cite{loeliger_2009}, making it likely to occur in LLM
pre-training data and thus giving LLMs some basic familiarity with it.
In this example, the edit has two search-replace blocks, one for each
of the two files given in the localization~(\Cref{lst:localization}).
In general, the tool also supports multiple search-replace blocks in
the same file.
Each search-replace block has two parts separated by
`\texttt{\small =======}': a block of lines to search in~$c_\textrm{old}$,
and a block of lines to replace them with to obtain~$c_\textrm{new}$.
The initial LLM context has few-shot samples for this format.

The edit agent tool performs a series of escalating checks on the LLM-generated
tool call, such that the first check that fails causes an early exit
with a corresponding observation to explain what went wrong.
It checks adherance to the merge-conflict format and attempts to match
the search blocks, with heuristic rule-based repairs to tolerate minor
differences such as including the agent-injected edit markers.
Once these initial checks pass, it derives the candidate~$c_\textrm{new}$
to be subjected to the remaining checks.
It runs a simple Java linter on~$c_\textrm{new}$ that does not have
side-effects but misses many deeper Java code problems.
Finally, only when all previous checks succeeded, it fires up a
containerized environment with the dependencies of the current
repository installed and kicks off the Java compiler on~$c_\textrm{new}$.
The containerization isolates the versions of Java tools and libraries from
the host machine and prevents any side-effects of the build scripts
from affecting the master copy of the code.
If all checks succeed without finding mistakes, the agent loop exits.

The above description applies to the editing agent in its default
configuration.
The agent can also be configured to elide certain information from
the prompt, such as the thought from the localization agent or the
proposed edit.
We can change the number of few-shot samples for the merge-conflict
format.
Furthermore, we can configure whether or not to run the Java
compiler using the project build scripts.

\section{Evaluation}\label{sec:evaluation}

\begin{table}[t]
\centering
\caption{Main results on Multi-SWE-Bench (Java) and SWE-PolyBench (Java).\protect\footnotemark[1]\protect\footnotemark[2]}
% \footnote{Correct Location metrics are unavailable for a subset of Multi-SWE-Bench (Java) leaderboard submissions due to issues encountered when applying the publicly available output patches.}\footnote{Cost and average token metrics are unavailable for SWE-PolyBench (Java) leaderboard submissions due to the lack of publicly available output patches.}

\label{tab:main_results}
\resizebox{\textwidth}{!}{
\begin{tabular}{c|l|rrr|rr|rr}
\toprule
 \multirow{3}{*}{\makecell{Bench\\mark}} & \multirow{3}{*}{Base LLM} & \multirow{3}{*}{\makecell{\% \\Resolved}} & \multirow{3}{*}{\makecell{\$ Cost \\ Total (Avg.)}} & \multirow{3}{*}{\makecell{Avg.\\ \# Tokens}} & \multicolumn{4}{c}{Edit Correct Location} \\
& & & & & \multicolumn{2}{c|}{File} & \multicolumn{2}{c}{Node}\\
& & & & & Recall & Precision & Recall & Precision\\
\midrule
\multicolumn{9}{c}{\textbf{iSWE-Agent (Frontier Models)}} \\
\cline{2-9}
\multirow{17}{*}{\makecell{\protect\rotatebox{90}{Multi-SWE-Bench (Java)}}}
& \textbf{Claude-4.5-Sonnet} & 43 (33.6\%) & \textbf{237.89 (1.86)} & 598k & 62\% & 82\% & 44\% & 68\% \\
& \textbf{Claude-3.7-Sonnet} & 32 (25.0\%) &  \textbf{154.51 (1.21)} & 386k & 50\% & 72\%& 32\% & 63\%\\
\cline{2-9}
\multicolumn{9}{c}{\textbf{Leaderboard submissions} (top other agents for comparison)} \\
\cline{2-9}
& \textit{InfCode (GPT-5.2)} & 50 (39.1\%) & \textbf{569.29 (4.78)} & 2453k & 54\% &72\%& 47\% &54\%\\
& \textit{MSWE-agent (C-3.7-Sonnet)$^1$} & 30 (23.4\%) & \textbf{477.26 (3.72)} & 1233k & - & - & - & - \\
& \textit{MopenHands (C-3.7-Sonnet)$^1$} & 28 (21.8\%) & \textbf{332.61 (2.59)} & 835k & - & - & - & - \\
% \textit{C-3.7-S MopenHands 144 instances} & 28 (21.88\%) & \textbf{381.25(2.65)} & 851k & -\% &-\%& -\% &-\%\\

\cline{2-9}
\multicolumn{9}{c}{\textbf{iSWE-Agent (Open Source Models)}} \\
\cline{2-9}

& deepSeek-V3-2               & 28 (21.8\%) & 32.46 (0.25) & 899k & 44\% & 57\% & 32\% & 48\% \\
& llama-4-maverick-17b-128e   & 21 (16.4\%) & 14.53 (0.11) & 82k & 37\% & 59\% & 22\% & 51\% \\
& gpt-oss-120b (tool calling) & 21 (16.4\%) &  4.86 (0.04) & 223k & 36\% & 57\% & 19\% & 47\% \\
% & DeepSeek-V3                 & 19 (14.84\%) & - & - & 47\% & 71\% & 19\% & 45\% \\
& deepSeek-R1                 & 19 (14.8\%) &  7.70 (0.06)  & 83k & 38\% & 54\% & 20\% & 42\% \\
& devstral-small-2505         & 19 (14.8\%) & 10.16 (0.08) & 775k & 40\% & 60\% & 24\% & 47\% \\
& qwen2-5-72b-instruct        & 17 (13.2\%) &  1.49 (0.01) & 89k & 34\% & 52\% & 20\% & 41\% \\
& llama-3.3-70b               & 15 (11.7\%) & 10.60 (0.08) & 116k & 35\% & 53\% & 20\% & 43\% \\
& devstral-small-2507         & 13 (10.1\%) & 15.98 (0.12) & 1235k & 30\% & 47\% & 18\% & 37\% \\
& gpt-oss-20b (tool calling)  & 11 ~(8.5\%) & 13.56 (0.11) & 1461k & 24\% & 37\% & 17\% & 33\% \\

\midrule
\multicolumn{9}{c}{\textbf{iSWE-Agent (Frontier Models)}} \\
\cline{2-9}
% -           &  0 (0.00\%)  & - & - &  -\% &  -\% &  -\% &  -\% \\
\multirow{13}{*}{\makecell{\protect\rotatebox{90}{SWE-PolyBench (Java)}}}
& Claude-4.5-Opus   &  55 (33.3\%) &  420.7 (2.55) & 492k & 61\% & 85\% & 44\% & 74\%\\
& Claude-4.5-Sonnet &  48 (29.1\%) &  305.13 (1.85) & 588k & 53\% & 72\% & 34\% & 63\%\\
\cline{2-9}
\multicolumn{9}{c}{\textbf{Leaderboard submissions} (top other agents for comparison)} \\
\cline{2-9}
& \textit{Amazon Q Developer Agent$^2$} & 44 (26.6\%) & - & - & 56\% & 75\%& 43\% & 62\%\\
& \textit{Aider-PB (C-3.5-Sonnet)$^2$}    & 26 (15.7\%) & - & - & 52\% & 58\%& 32\% & 52\%\\
\cline{2-9}

\multicolumn{9}{c}{\textbf{iSWE-Agent (Open Source Models)}}\\
\cline{2-9}

& deepSeek-V3-2               & 30 (18.1\%) & 46.5 (0.28) & 1000k & 35\% & 45\% & 23\% & 40\% \\
& gpt-oss-120b (tool calling) & 23 (13.9\%) & 8.01 (0.05) & 290k & 27\% & 44\% & 14\% & 38\% \\
& deepSeek-R1                 & 20 (12.1\%) & 10.64 (0.06) & 62k & 34\% & 49\% & 19\% & 42\% \\
& llama-4-maverick-17b-128e   & 17 (10.3\%) & 20.71 (0.13) & 91k & 33\% & 50\% & 18\% & 42\% \\
& qwen2-5-72b-instruct        & 14 ~(8.4\%) &  2.15 (0.01) & 100k & 26\% & 39\% & 13\% & 33\% \\
& devstral-small-2505         & 14 ~(8.4\%) &  8.76 (0.05) & 522k & 28\% & 39\% & 14\% & 31\% \\
& devstral-small-2507         & 13 ~(7.8\%) & 17.52 (0.11) & 1051k & 28\% & 39\% & 14\% & 31\% \\
& gpt-oss-20b (tool calling)  & 10 ~(6.0\%)  & 17.54 (0.11) & 1464k & 13\% & 24\% & 7\% & 20\% \\
& llama-3.3-70b               & ~9 ~(5.4\%) & 16.51 (0.10) & 140k & 23\% & 36\% & 12\% & 29\% \\

% -  &  0 (0.00\%) &  - & - & -\% & -\% & -\% & -\%\\

\bottomrule
\end{tabular}
}
\end{table}

\footnotetext[1]{Correct Location metrics are unavailable for a subset of Multi-SWE-Bench (Java) leaderboard submissions due to difficulties encountered when applying the publicly available output patches.}

\footnotetext[2]{Cost and average token metrics are unavailable for SWE-PolyBench (Java) leaderboard submissions due to the lack of publicly available trajectories.}

\begin{figure}[t]
  \includegraphics[width=0.499\textwidth]{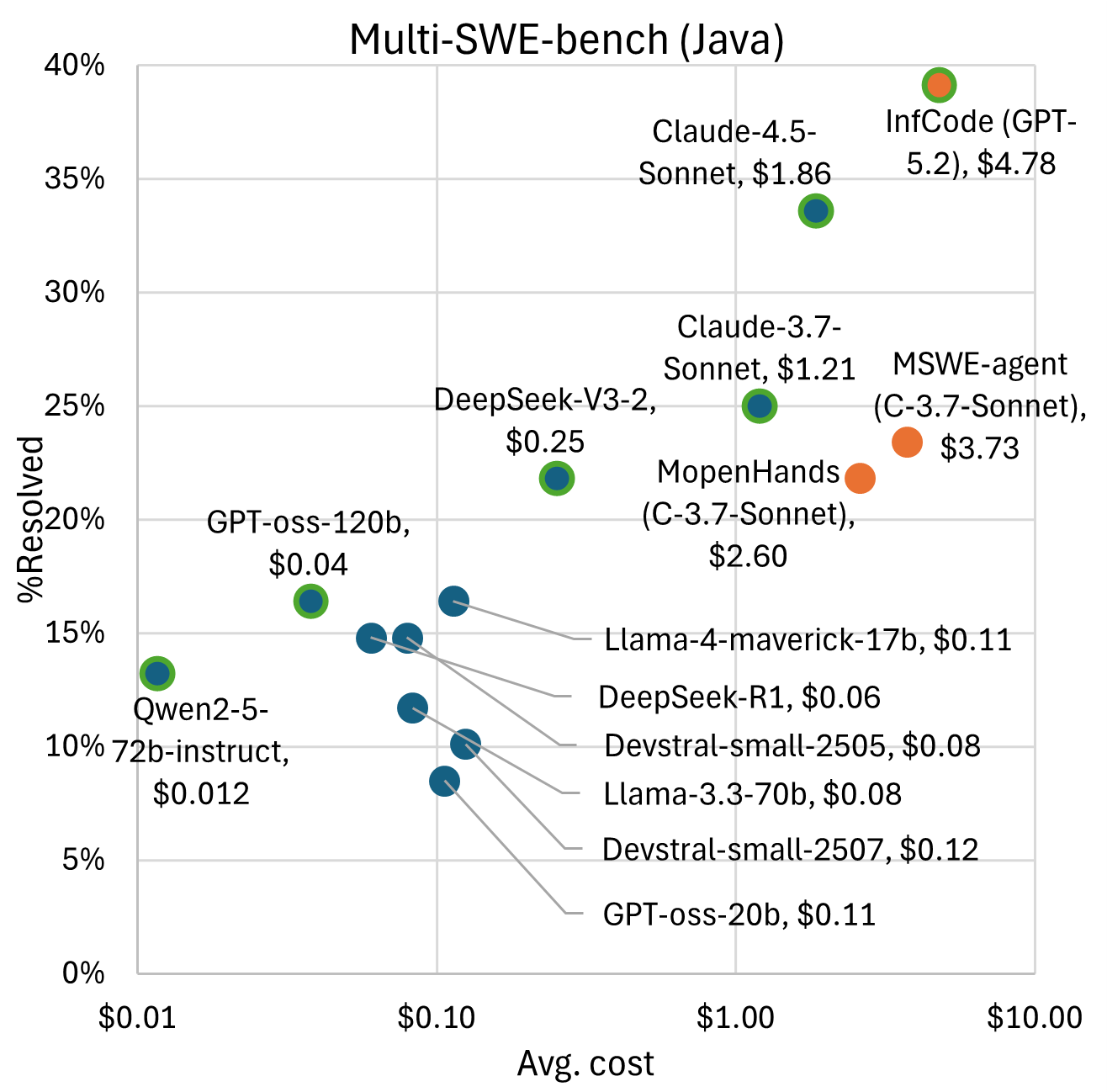}
  \includegraphics[width=0.499\textwidth]{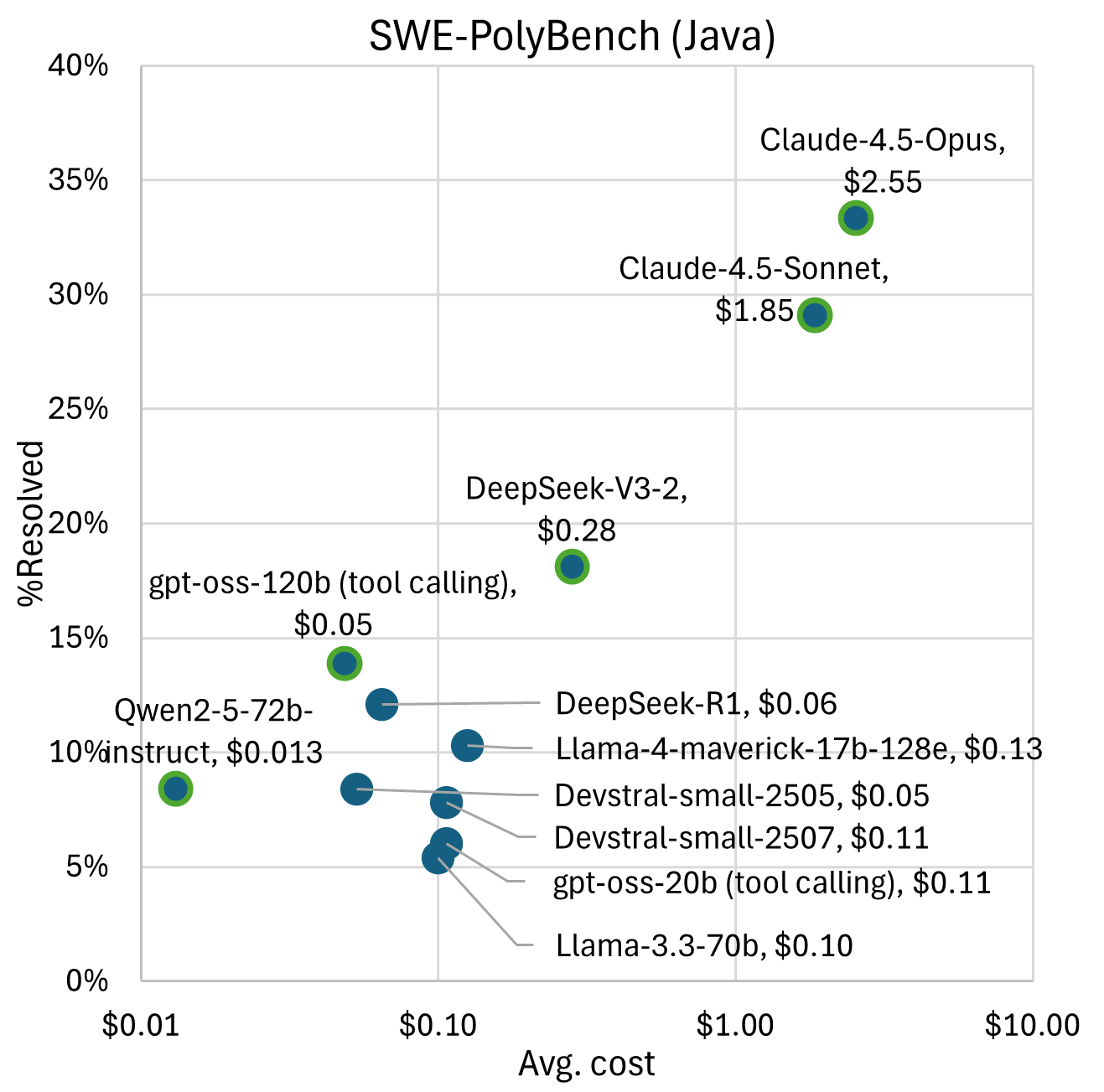}
  \caption{\label{fig:cost_v_resolved_models}Average cost vs.\ \%resolved across models. Most datapoints are from iSWE with different models; two datapoints for Multi-SWE-bench (Java) are from MopenHands and MSWE-agent.}
\end{figure}

This section presents our empirical evaluation of iSWE based on the
Java splits of \emph{Multi-SWE-bench}~\cite{zan_et_al_2025} and
\emph{SWE-PolyBench}~\cite{rashid_et_al_2025}.
Since there has been little published work on using those benchmarks
so far, this sheds light not just on iSWE's performance but also on
the benchmarks themselves.
This section starts by discussing the main results in
\Cref{tab:main_results}, and then dives deeper into various aspects
with additional tables and figures.
\Cref{tab:main_results} is based on running iSWE on the two benchmarks
with a variety of different LLMs.
We use inline tool calling for most models except for gpt-oss, where
localization used native tool calling, because that performed better.
For comparison, we also included
some results from other top agents on the corresponding leaderboards.

Column \textbf{\% Resolved} indicates for how many benchmark instances the
agent-generated patch~$c_\textrm{new}$ resolves the issue~$d_\textrm{issue}$
as indicated by passing the hidden tests~$t_\textrm{gold}$.
With frontier models, iSWE has among the highest resolution rates
on Java issues.
While the open-source models resolve fewer instances, they also show
descent results, demonstrating iSWE's ability to use different models.
The absolute numbers are lower than commonly seen on Python
leaderboards; this is likely an artifact of Python leaderboards
becoming saturated~\cite{openai_verified_no_longer_2026}.
Column \textbf{\$ Cost} shows the expense for LLM calls from model
vendor APIs in US dollars.
To calculate it, we track the number of input and output tokens, and
then use public pricing per input or output token, mostly based on
scaling factors included in the LiteLLM library.
Generally, stronger models cost more.
For Multi-SWE-bench, the top non-iSWE results use Claude-3.7-Sonnet;
the iSWE runs with the same model incur 2$\times$ to 3$\times$ lower
cost while resolving more issues.
Column \textbf{Avg.\ \# Tokens} shows the average number of tokens
used per instance.
It varies across models with no clear correlation to resolution rate.
Columns \textbf{Edit Correct Location} show to what extent the
agent-generated new code~$c_\textrm{new}$ modifies the same code
locations as the hidden golden patch.
The columns report precision and recall both at the file level and at
the level of \emph{nodes}, which are the closest enclosing class or
function that contains a given diff-hunk~\cite{rashid_et_al_2025}.
Not surprisingly, file-level retrieval numbers are higher than
node-level, and retrieval metrics correlate well with \% resolved.

The following subsections explore iSWE and the benchmarks more deeply.
\Cref{sec:results_cost} explores dollar costs and number of turns by
model and instance to better highlight the trade-offs enabled by iSWE.
\Cref{sec:results_subagents} digs deeper into the sub-agent and even
tool layer of iSWE to explore how each component contributes to
overall success.
And \Cref{sec:results_samples} provides a better understanding of the
benchmarks by drilling down to samples of different complexities and
other characteristics.

\subsection{Results for Cost and Turns}\label{sec:results_cost}

While the primary metric for SWE-bench and similar benchmarks is
issue resolution rate, another major concern is cost.
Furthermore, one can trade resolution rate and cost
off against each other, e.g., by choosing a different
model or by resorting to inference scaling.
This subsection explores that trade-off in more detail, using dollar
costs computed based on published pricing for LLM input and output
tokens~(mostly leveraging metadata in LiteLLM).
Besides dollar cost, this also sheds light on other correlated
concerns, such as the number of turns in the agentic loop or the
time-to-fix.
All results in this section are based on the same runs as the main
results.

\Cref{fig:cost_v_resolved_models} visualizes the trade-off between
\mbox{\% resolved} and average per-instance \mbox{\$ cost} as a
scatter plot.
The x-axis shows cost on a logarithmic scale; cheaper (further left)
is better.
The y-axis shows resolution rate on a linear scale; more resolved
(further up) is better.
The points on the Pareto frontier are those on the upper left and are
highlighted with a green border: they are partially optimal in the
sense that other points may be higher in one metric but never both.
For both benchmarks, the Pareto frontier is occupied by iSWE with five
models: qwen2-5-72b-instruct, gpt-oss-120b, deepseek-v3-2, and finally
two versions of the Claude frontier models.
Thanks to iSWE being model agnostic, it offers the user a range of
choices to navigate the cost/accuracy trade-off.
When we first did these experiments, the top non-iSWE agents on
Multi-SWE-bench~(Java) were MSWE-agent and MopenHands, both using Claude-3.7-Sonnet.
They are dominated by iSWE with the same model, which achieves a
higher resolution rate at a fraction of the cost.
In the meantime, InfCode using GPT-5.2 has also appeared on the
Multi-SWE-bench~(Java) leaderboard, with higher resolved rate but at
higher cost than iSWE.

\begin{figure}[t]
    \centering
    % Top Row
    \begin{subfigure}[b]{0.45\textwidth}
        \centering
        \includegraphics[width=\textwidth]{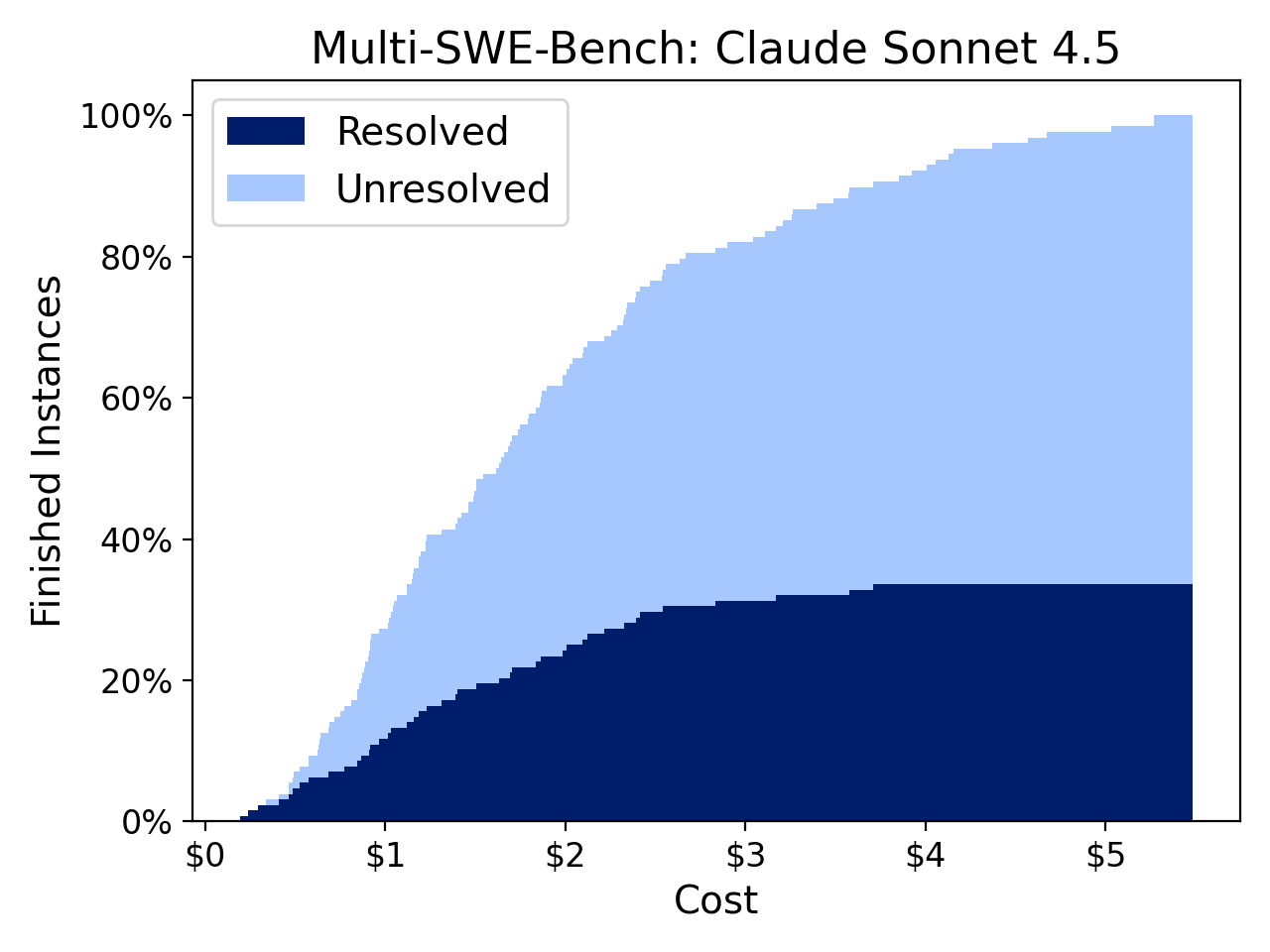}
        % \caption{}
        \label{fig:a}
    \end{subfigure}
    \begin{subfigure}[b]{0.45\textwidth}
        \centering
        \includegraphics[width=\textwidth]{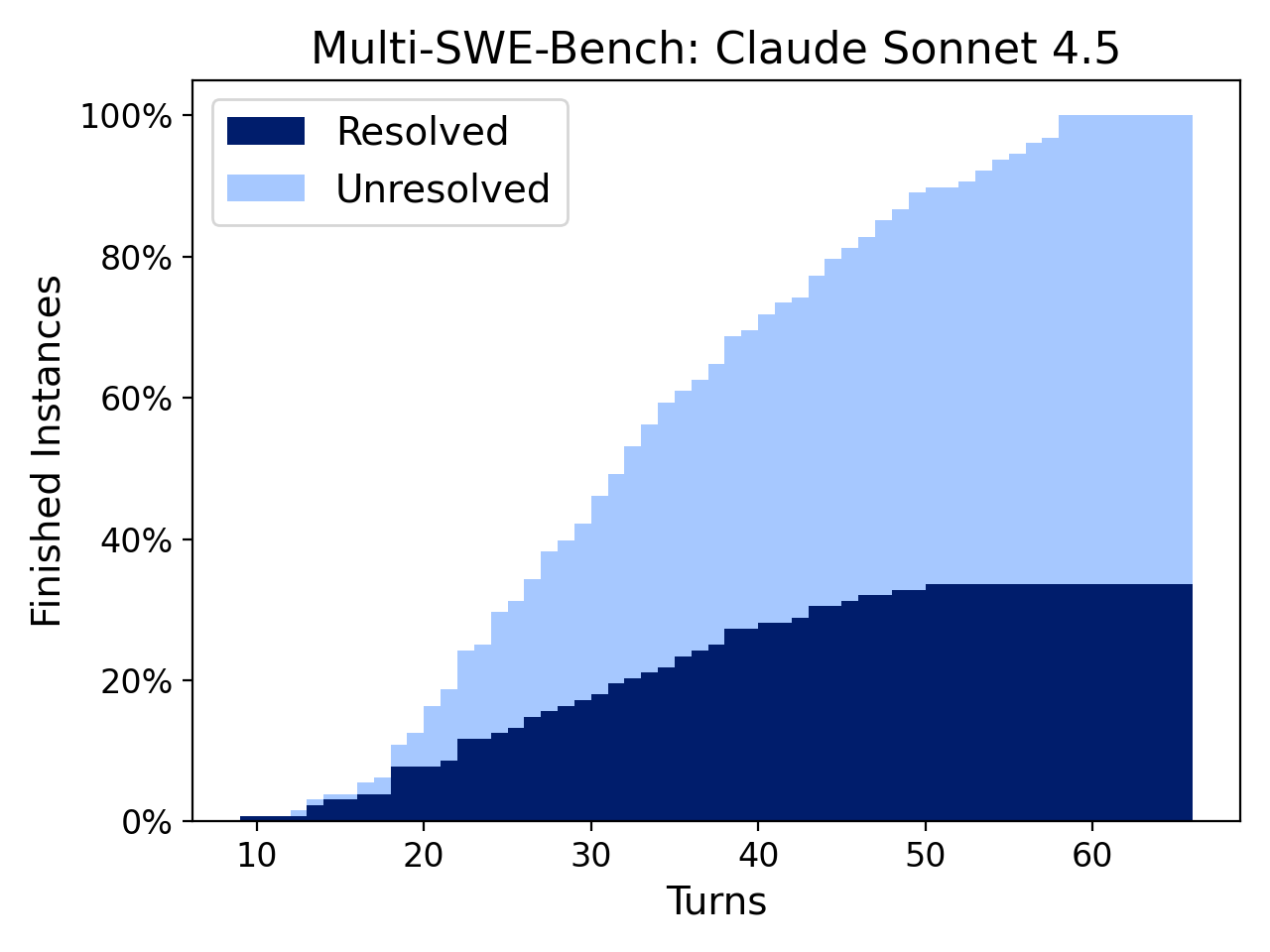}
        % \caption{}
        \label{fig:b}
    \end{subfigure}
    \\[-2mm]
    % Bottom Row
    \begin{subfigure}[b]{0.45\textwidth}
        \centering
        \includegraphics[width=\textwidth]{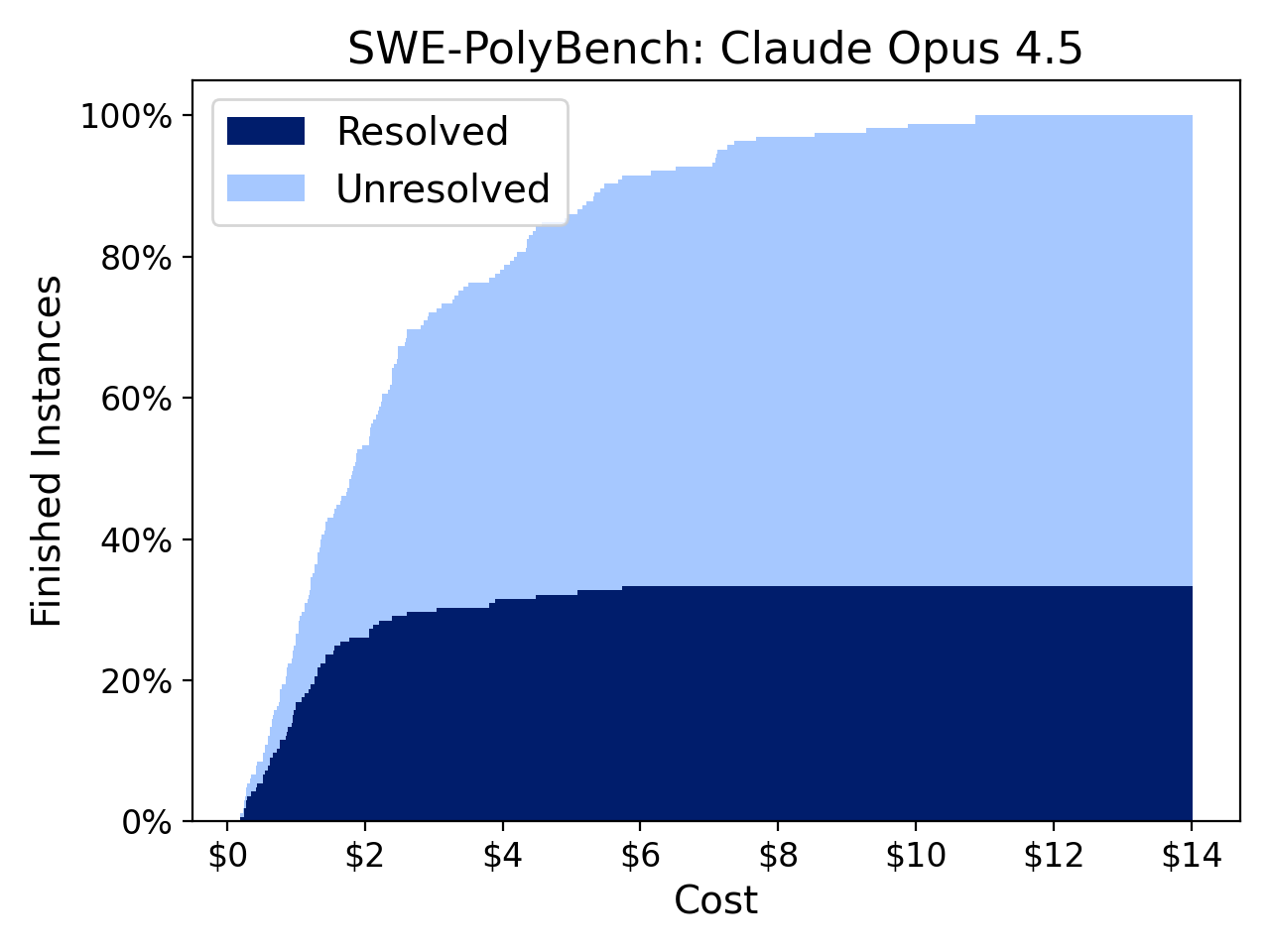}
        % \caption{}
        \label{fig:c}
    \end{subfigure}
    \begin{subfigure}[b]{0.45\textwidth}
        \centering
        \includegraphics[width=\textwidth]{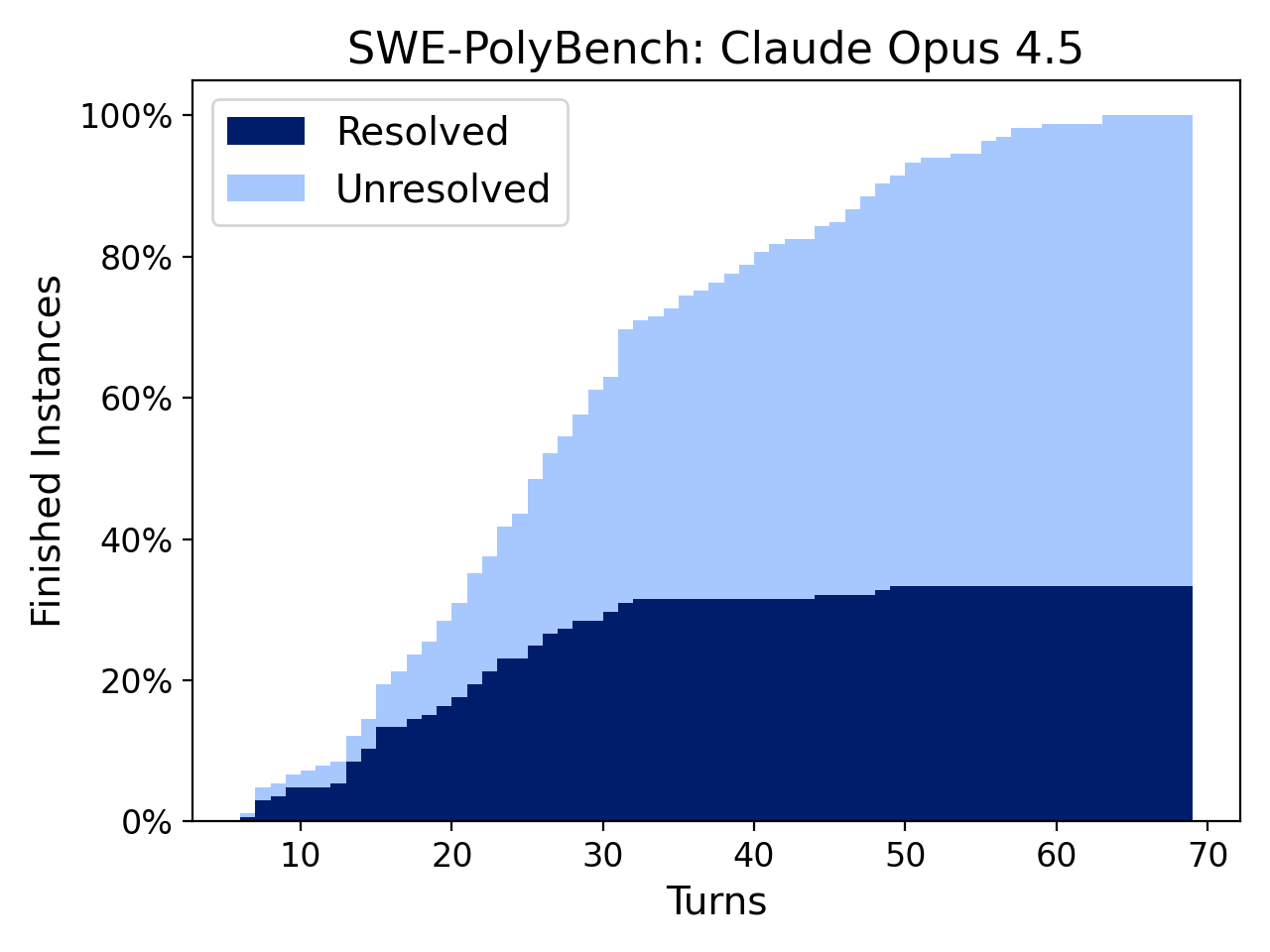}
        % \caption{}
        \label{fig:d}
    \end{subfigure}
    \\[-5mm]
    \caption{\label{fig:cost_v_resolved_instances}Average cost vs.\ resolved across individual instances in the benchmark (cumulative).}
\end{figure}

While the previous figure showed average costs across all instances in
an entire benchmark, not all instances incur the same cost.
Some instances can be resolved in fewer turns or tokens, and thus at
lower cost, than others.
\Cref{fig:cost_v_resolved_instances} shows the cumulative distribution
of how many instances get finished~(light blue) or resolved~(dark
blue) up to a cost (or turns) threshold shown on the x-axis.
The left two subfigures focus on dollar cost: most instances that get
resolved take at most~\$4.
The number of finished instances keeps increasing beyond that point,
but they experience diminishing returns in terms of resolution.
The right two subfigures have the same y-axis but put the number of
turns on the x-axis.
Early turns are cheaper, because as context accumulates, later turns
consume more tokens.
Again, the curve for resolved issues asymptotically flattens at fewer
turns than the curve for finished instances, indicating that LLMs
struggle with long-horizon tasks.
These results indicate that cost thresholds are another lever
for navigating the cost/accuracy trade-off.

\begin{table}[t]
\centering
\caption{Cost per sub-agent.}
\label{tab:cost_per_sub_agent}
\resizebox{0.9\textwidth}{!}{
\begin{tabular}{c|l|r|r|r|r|r|r}
\toprule
\multirow{2}{*}{\makecell{Benchmark}} & \multirow{2}{*}{Base LLM} &  \multicolumn{3}{c|}{Avg.\ \# Turns} & \multicolumn{3}{c}{Avg.\ Cost} \\
\cmidrule(lr){3-5}
\cmidrule(lr){6-8}
& & Loc. & Editing & Total & Loc. & Editing & Total \\
\midrule
\multirow{2}{*}{Multi-SWE-\newline{}Bench (Java)}
& Claude-4.5-Sonnet (Loc \& Edit) & 31.5 & 1.7 & 33.3 & \$1.77 & \$0.09 & \$1.86 \\
& DeepSeek-V3-2 (Loc \& Edit) & 41.6 & 2.5 & 44.2 & \$0.24 & \$0.01 & \$0.25 \\
\midrule
\multirow{2}{*}{SWE-PolyBench (Java)}
& Claude-4.5-Opus (Loc \& Edit) & 34.1 & 2.2 & 36.3 & \$2.41 & \$0.14 & \$2.55 \\
& DeepSeek-V3-2 (Loc \& Edit) & 54.8 & 3.3 & 58.0 & \$0.27 & \$0.01 & \$0.28 \\
\bottomrule
\end{tabular}
}
\end{table}

\Cref{tab:cost_per_sub_agent} breaks down the cost spent in
iSWE's two sub-agents for localization and editing, respectively.
For both benchmarks, for two representative models from the Pareto
frontier, most of the turns and cost are incurred during localization.
Even though DeepSeek takes more turns than Claude, in the end it
incurs lower cost, because its per-token pricing is cheaper.
Note that iSWE allows different models for localization and editing.
For example, we did a run on Multi-SWE-bench~(Java) where we used
Claude-4.5-Sonnet for localization and Gemini-2.5-Pro for editing~(elided from \Cref{tab:main_results}).
That run resolved 29.6\% of instances at an average cost of \$2.02 per
instance.

% =============================================================

\subsection{Results by iSWE Sub-Agents and Tools}\label{sec:results_subagents}

\begin{table}[t]
\centering
\caption{Localization metrics per sub-agent.}
\label{tab:localization_per_sub_agent}
\resizebox{\textwidth}{!}{
\begin{tabular}{c|l|cc|cc|cc|cc}
\toprule
\multirow{2}{*}{Benchmark} & \multirow{2}{*}{Base LLM}
& \multicolumn{4}{c|}{Localization Correct Location}
& \multicolumn{4}{c}{Edit Correct Location} \\
& 
& \multicolumn{2}{c|}{File}
& \multicolumn{2}{c|}{Node}
& \multicolumn{2}{c|}{File}
& \multicolumn{2}{c}{Node} \\
& 
& Recall & Precision & Recall & Precision
& Recall & Precision & Recall & Precision \\
\midrule

\multicolumn{10}{c}{\textbf{iSWE-Agent (Frontier Models)}} \\
\cline{2-10}
\multirow{14}{*}{\protect\rotatebox{90}{Multi-SWE-Bench (Java)}}
& Claude-4.5-Sonnet & 63\% & 83\% & 36\% & 59\% & 62\% & 82\% & 44\% & 68\% \\
% & C-4.5-S (Loc), G-2.5-Pro (Edit) & 63\% & 83\% & 36\% & 59\% & 54\% & 74\% & 38\% & 61\% \\
& Claude-3.7-Sonnet & 56\% & 77\% & 28\% & 53\% & 50\% & 72\% & 32\% & 63\% \\

\cline{2-10}
\multicolumn{10}{c}{\textbf{iSWE-Agent (Open Source Models)}} \\
\cline{2-10}
& deepSeek-V3-2 & 58\% & 76\% & 34\% & 55\% & 44\% & 57\% & 32\% & 48\%  \\
& llama-4-maverick-17b-128e & 45\% & 70\% & 21\% & 48\% & 37\% & 59\% & 22\% & 51\% \\
& gpt-oss-120b (tool calling) & 49\% & 75\% & 20\% & 48\% & 36\% & 57\% & 19\% & 47\%  \\
& deepSeek-R1 & 49\% & 72\% & 21\% & 46\% & 38\% & 54\% & 20\% & 42\% \\
& devstral-small-2505 & 44\% & 67\% & 22\% & 44\% & 40\% & 60\% & 24\% & 47\%  \\
& qwen2-5-72b-instruct & 44\% & 63\% & 18\% & 40\% & 34\% & 52\% & 20\% & 41\% \\
& llama-3.3-70b & 40\% & 60\% & 17\% & 39\% & 35\% & 53\% & 20\% & 43\% \\
& devstral-small-2507 & 36\% & 53\% & 17\% & 35\% & 30\% & 47\% & 18\% & 37\% \\
& gpt-oss-20b (tool calling) & 36\% & 55\% & 22\% & 40\% & 24\% & 37\% & 17\% & 33\% \\

\midrule
\multicolumn{10}{c}{\textbf{iSWE-Agent (Frontier Models)}} \\
\cline{2-10}
\multirow{14}{*}{\protect\rotatebox{90}{SWE-PolyBench (Java)}}
& Claude-4.5-Opus & 62\% & 83\% & 31\% & 60\% & 56\%& 79\% & 38\% & 72\% \\
& Claude-4.5-Sonnet & 61\% & 80\% & 29\% & 56\% & 53\% & 72\% & 34\% & 63\% \\
\cline{2-10}

\multicolumn{10}{c}{\textbf{iSWE-Agent (Open Source Models)}} \\
\cline{2-10}
& deepSeek-V3-2 & 56\% & 73\% & 28\% & 52\% & 35\% & 45\% & 23\% & 40\%  \\
& gpt-oss-120b (tool calling) & 45\% & 69\% & 19\% & 50\% & 27\% & 44\% & 14\% & 38\%  \\
& deepSeek-R1 & 48\% & 69\% & 18\% & 42\% &  34\% & 49\% & 19\% & 42\%  \\
& llama-4-maverick-17b-128e & 46\% & 70\% & 17\% & 45\% &  33\% & 50\% & 18\% & 42\%  \\
& qwen2-5-72b-instruct & 38\% & 58\% & 14\% & 39\% & 26\% & 39\% & 13\% & 33\%  \\
& devstral-small-2505 & 39\% & 56\% & 15\% & 36\% &  28\% & 39\% & 14\% & 31\%  \\
& devstral-small-2507 & 36\% & 52\% & 15\% & 34\% &  28\% & 39\% & 14\% & 31\%  \\
& gpt-oss-20b (tool calling) & 36\% & 52\% & 14\% & 35\% & 13\% & 24\% & 7\% & 20\%  \\
& llama-3.3-70b & 39\% & 57\% & 15\% & 33\% & 23\% & 36\% & 12\% & 29\%  \\

\bottomrule
\end{tabular}
}
\end{table}

% dollar cost (!!!)
% Cost analysis of Loc/ Edit agents based on tokens consumption #619 @shinnar WIP
% localization metrics
% Loc Agent metrics calculation #617 pt. (2) - Check whether our node definition aligns with SWE-PolyBench @NEVO - Pending
% tool distribution
% Edit agent - tool distribution #654 @Sami-Serhan
% number of turns (in localization and editing) - already saved in trajectory
% lower priority:
% end-to-end time - Skip for now
% number of diff hunks
% Edit agent - tool distribution #654 @Sami-Serhan
% number of non-Java edits Edit agent - tool distribution #654 @Sami-Serhan

% - by complexity,
%   * Easy, medium, Hard - MSB
%   * is_single_class, is_single_fn etc for SWE-PB
% - by issue date (for contamination)
% - by which SWE-PolyBench instances are in the "Verified" subset
% - by problems of Multi-SWE-bench
%   * unresolved with gold patch
%   * duplicates
%   * hints provided

\begin{table}[t]
\centering
\caption{Tool usage distribution across the entire benchmark~(not per-instance).}
\label{tab:tool_usage_distribution}
\resizebox{\textwidth}{!}{
\begin{tabular}{c|l|rrrr|rrr|r}
\toprule
\multirow{2}{*}{\makecell{Bench\\mark}} &
\multirow{2}{*}{Base LLM} &
\multicolumn{7}{c|}{Tool Usage (Number of Calls)} &
\multirow{2}{*}{Total} \\
\cmidrule(lr){3-9}
 & &
\makecell{get\\file\\info} &
\makecell{get\\class\\info} &
\makecell{get\\method\\info} &
\makecell{get\\symbol\\info} &
\makecell{get\\inheritance\\hierarchy} &
\makecell{get\\function\\callers} &
\makecell{get\\call\\chain} &
\\
\midrule

\multicolumn{10}{c}{\textbf{Frontier Models}} \\
\cline{2-10}
\multirow{10}{*}{\makecell{\protect\rotatebox{90}{Multi-SWE-Bench (Java)}}}
& Claude-4.5-Sonnet           &   87 & 124 & 403 & 181 & 2 & 19 & 33 & 849 \\
& Claude-3.7-Sonnet           &  163 & 136 & 265 &  75 & 1 &  7 &  5 & 652 \\

\cline{2-10}
\multicolumn{10}{c}{\textbf{Open Source Models}} \\
\cline{2-10}
& DeepSeek-V3-2               &  497 & 549 & 1803 & 1113 & 22 & 93 &  49 & 4126 \\
& llama-4-maverick-17b-128e   &   73 & 245 &  357 &   90 &  2 &  7 &  30 &  804 \\
& gpt-oss-120b (tool calling) &  604 & 220 &  708 &  224 &  0 &  1 &   1 & 1758 \\
& Devstral-small-2505         & 2207 & 478 & 1431 &  408 & 29 & 20 &  15 & 4588 \\
& Qwen2-5-72b-instruct        &  134 & 389 &  696 &   84 &  5 &  2 &  59 & 1369 \\
& llama-3.3-70b               &   65 & 307 &  508 &   55 &  7 & 38 & 101 & 1081 \\
& Devstral-small-2507         & 1933 & 384 & 3512 &  360 & 11 & 20 &  12 & 6232 \\
& gpt-oss-20b (tool calling)  & 4076 & 178 & 1239 &  281 &  0 &  5 &  13 & 5792 \\

\midrule
\multicolumn{10}{c}{\textbf{Frontier Models}} \\
\cline{2-10}
\multirow{13}{*}{\makecell{\protect\rotatebox{90}{SWE-PolyBench (Java)}}}
& Claude-4.5-Opus             &  416 & 734 & 1985 &  848 & 39 & 82 &  96 & 4200 \\
& Claude-4.5-Sonnet           &  127 & 184 &  466 &  226 &  2 & 33 &  41 & 1079 \\

\cline{2-10}
\multicolumn{10}{c}{\textbf{Open Source Models}} \\
\cline{2-10}
& DeepSeek-V3-2               &  691 &  878 & 2397 & 1653 & 21 & 114 &  73 & 5827 \\
& gpt-oss-120b (tool calling) &  929 &  438 &  999 &  422 &  0 &   3 &   0 & 2791 \\
& DeepSeek-R1                 &   66 &  342 &  549 &  106 &  8 &  17 &  22 & 1110 \\
& llama-4-maverick-17b-128e   &   83 &  393 &  459 &  180 &  6 &   8 &  28 & 1157 \\
& Qwen2-5-72b-instruct        &  291 &  656 &  779 &  164 &  2 &  12 &  24 & 1928 \\
& Devstral-small-2505         & 2019 &  942 & 2126 &  631 &  9 &  29 &   5 & 5761 \\
& Devstral-small-2507         & 2171 & 1203 & 4213 &  954 &  7 &  33 &   7 & 8588 \\
& gpt-oss-20b (tool calling)  & 7545 &  236 & 1075 &  604 &  0 &   0 &  37 & 9497 \\
& llama-3.3-70b               &   58 &  369 &  702 &   86 & 18 &  82 & 124 & 1439 \\

\bottomrule
\end{tabular}
}
\end{table}

\begin{table}[t]
\centering
\caption{Localization performance for different tool sets, using Claude-4.5-Sonnet on Multi-SWE-bench~(Java).}
\label{tab:different_tool_sets}
\resizebox{\textwidth}{!}{
\begin{tabular}{l|r|r|r|rr|rr}

\toprule
\multirow{2}{*}{Tool Set} & \multirow{2}{*}{\$ Cost} & \multirow{2}{*}{Avg.} & \multirow{2}{*}{Turns} & \multicolumn{2}{c|}{File Retrieval} & \multicolumn{2}{c}{Node Retrieval} \\
\cmidrule(lr){5-6} \cmidrule(lr){7-8}
 & Total (Avg.) & \#Tokens & & Recall & Precision & Recall & Precision \\
\midrule
iSWE-tools & 226 (1.77) & 575k & 4037 & 63\% & 83\% & 36\% & 59\% \\
iSWE-tools \& view-file & 229 (1.79) & 584k & 3777 & 63\% & 83\% &37\% & 57\% \\
Bash     & 263 (2.06) & 667k & 5815 &62\% & 82\% & 37\% & 57\% \\
All-Tools (iSWE-tools \& Bash \& swea) & 336 (2.63) & 855k & 5820 &  62\% & 81\% & 37\% & 58\% \\

\bottomrule
\end{tabular}
}
\end{table}

This subsection breaks down how each component of iSWE
contributes to its overall success.

\Cref{tab:localization_per_sub_agent} shows the localization metrics
at two stages in the pipeline: 
based on the intermediate JSON right after the localization
agent~(see e.g.\ \Cref{lst:localization}),
and based on the final code patch~$c_\textrm{new}$ after the editing
agent~(see e.g.\ \Cref{lst:edit}).
Calculating the same metrics at these different stages sheds light on
the quality of the localization-agent intermediate output as well as
on the faithfulness of the editing agent in following it.
As before, the node metrics are based on the nearest enclosing
function or class, similar to SWE-PolyBench~\cite{rashid_et_al_2025}.
\Cref{tab:localization_per_sub_agent} shows that file-level metrics
are better after localization than after edit, because iSWE only
shows the files found by the localization to the edit agent.
In contrast, node-level metrics are sometimes better after edit than after
localization, because the edit agent can further adjust its location.
As we saw before, localization metrics closely track resolution rate,
indicating that good localization is important to overall success.

\Cref{tab:tool_usage_distribution} shows how often the various models
call the various localization tools.
The first four tools~(\textit{get\_file/class/method/symbol\_info})
answer questions about code-graph nodes, whereas the last three
tools~(\textit{get\_inheritance\_hierarchy/function\_callers/call\_chains})
answer questions about edges.
There are more calls to node tools than to edge tools; one reason for this
may be that edge tools yield more useful information in a single call.
The overall most-called tool is \textit{get\_method\_info} and the
least-call tool is \textit{get\_inheritance\_hierarchy}.
Aside from a few exceptions, all models call pretty much all tools, indicating that iSWE's tools are model-agnostic.

\Cref{tab:different_tool_sets} explores how iSWE's tools compare
to alternatives.
It is based on multiple different benchmark runs, each with a
different tool-set configuration for the localization agent.
Besides iSWE's Java-aware tools introduced
in \Cref{sec:approach}, alternative tools include
\textit{view\_file}; \textit{bash}~(running low-level shell commands);
and \textit{swea}~(the set of tools from SWE-agent~\cite{yang_et_al_2024},
which are read-only but lower-level than the iSWE tools).
The \textit{bash} tool, despite being low-level, is feasible with
recent frontier models, because those have been fine-tuned for SWE
tasks.
The iSWE tools, \textit{view\_file}, and the \textit{swea} tools are
read-only and free of side-effects;
in contrast, \textit{bash} can modify state and have other
potentially harmful side-effects, thus requiring sand-boxing.
Overall, the retrieval metrics are similar for all tool-sets.
On the other hand, cost and tokens are lower when using iSWE-tools,
through a combination of fewer turns and fewer spurious tokens per turn.
This is a symptom of the iSWE tools being higher-level: 
models find it easier to call them~(fewer turns)
and to interpret the output from calls~(fewer tokens).
Cost and retrieval metrics aside, using iSWE tools has the additional
advantage of avoiding spurious side-effects, making the agent safer
and simpler.

% =============================================================

\subsection{Results by Benchmark Sample Characteristics}\label{sec:results_samples}

\begin{table}[t]
\centering
\caption{Performance by hand-annotated complexity level on Multi-SWE-Bench (Java) and by hand-annotated Verified or not label on SWE-PolyBench (Java).}
\label{tab:complexity_hand_annotated}
\resizebox{\textwidth}{!}{
\begin{tabular}{c|l|c|l|r|c|cc|cc}
\toprule
\multirow{2}{*}{\makecell{Bench\\mark}} &
\multirow{2}{*}{Complexity} & \multirow{2}{*}{\makecell{\#\\Instances}} & \multirow{2}{*}{Base LLM} & \multirow{2}{*}{\% Resolved} & \multirow{2}{*}{Avg. Cost} & \multicolumn{2}{c|}{File Retrieval} & \multicolumn{2}{c}{Node Retrieval} \\
\cmidrule(lr){7-8} \cmidrule(lr){9-10}
& & & & & & Recall & Precision & Recall & Precision \\
\midrule
\multirow{6}{*}{\makecell{\rotatebox{90}{Multi-SWE-}}{\makecell{\rotatebox{90}{Bench (Java)}}}}
& \multirow{2}{*}{Easy} & \multirow{2}{*}{27} & Claude-4.5-Sonnet & 16 (59.2\%) & \$1.65 & 65\% & 80\% & 42\% & 49\% \\
 && & DeepSeek-V3-2 & 9 (33.3\%) & \$0.18 & 58\% & 74\% & 39\% & 47\% \\
 \cline{2-10}
 &\multirow{2}{*}{Medium} & \multirow{2}{*}{65} & Claude-4.5-Sonnet & 26 (40.0\%) & \$1.71 & 66\% & 86\% & 39\% & 58\%   \\
 && & DeepSeek-V3-2 & 18 (27.6\%) & \$0.21 & 61\% & 70\% & 37\% & 47\% \\
 \cline{2-10}
&\multirow{2}{*}{Hard} & \multirow{2}{*}{36} & Claude-4.5-Sonnet & 1~~(2.7\%) & \$2.26 & 55\% & 80\% & 22\% & 65\% \\
& & & DeepSeek-V3-2 & 1~~(2.7\%) & \$0.37 & 52\% & 87\% & 23\% & 75\% \\
\midrule\midrule
\rule{0pt}{14pt}\multirow{4}{*}{\makecell{\rotatebox{90}{SWE-Poly-}}{\makecell{\rotatebox{90}{Bench (Java)}}}}
&\multirow{2}{*}{Verified} & \multirow{2}{*}{69} & Claude-4.5-Opus & 31 (44.9\%) & \$2.25 & 67\% & 84\% & 33\% & 61\% \\
& & & DeepSeek-V3-2 & 18 (26.0\%) & \$0.28 & 60\% & 71\% & 30\% & 50\% \\
\cline{2-10}
\rule{0pt}{14pt}&\multirow{2}{*}{Non-Verified} & \multirow{2}{*}{96} & Claude-4.5-Opus & 25 (26.0\%) & \$2.76 & 58\% & 82\% & 28\% & 58\% \\
& & & DeepSeek-V3-2 & 12 (12.5\%) & \$0.27 & 53\% & 74\% & 26\% & 52\% \\
\bottomrule
\end{tabular}
}
\end{table}

\begin{table}[t]
\centering
\caption{Performance by localization-derived complexity attributes. These attributes are computed based on the gold patch and emulate the attributes defined by SWE-PolyBench.}
\label{tab:complexity_loc_derived}
\resizebox{\textwidth}{!}{
\begin{tabular}{c|l|c|l|r|r|rr|rr}
\toprule
\multirow{2}{*}{\makecell{Bench\\mark}} &
\multirow{2}{*}{\makecell{Modification\\Type}} & \multirow{2}{*}{\makecell{\#\\Instances}} & \multirow{2}{*}{Base LLM} & \multirow{2}{*}{\% Resolved} & \multirow{2}{*}{\makecell{Avg.\\Cost}} & \multicolumn{2}{c|}{File Retrieval} & \multicolumn{2}{c}{Node Retrieval} \\
\cmidrule(lr){7-8} \cmidrule(lr){9-10}
& & & & & & Recall & Precision & Recall & Precision \\
\midrule
\multirow{14}{*}{\makecell[c]{\rotatebox{90}{Multi-SWE-Bench (Java)}}}
& \multirow{2}{*}{None} & \multirow{2}{*}{3} & Claude-4.5-Sonnet & 0~~(0.0\%) & \$2.27 & 36\% & 100\% & 36\% & 100\%  \\
 && & DeepSeek-V3-2 & 0~~(0.0\%) & \$0.14 & 25\% & 100\% & 25\% & 100\% \\
 \cline{2-10}
 % &\multirow{4}{*}{\begin{tabular}{c|c}
 %      \multirow{4}{*}{\makecell[c]{Function Only}}&\multirow{2}{*}{\makecell[c]{Single}}  \\
 %      \cline{2-2}
 %      &\multirow{2}{*}{Total}  \\
 %      \end{tabular}}&  \multirow{2}{*}{??} & Claude-4.5-Sonnet & &  &  &  &  &    \\
 &\multirow{4}{*}{\makecell[c]{Function\\Only}\quad\makecell[c]{Single\\\\All}} &  \multirow{2}{*}{33} & Claude-4.5-Sonnet & 20 (60.6\%) & \$1.75 & 72\% & 87\% & 58\% & 70\%    \\
 && & DeepSeek-V3-2 & 10 (30.3\%) & \$0.23 & 56\% & 69\% & 50\% & 61\%  \\
 \cline{3-10}
&&\multirow{2}{*}{44} & Claude-4.5-Sonnet & 27 (61.3\%) & \$1.63 & 73\% & 90\% & 59\% & 74\%    \\
 && & DeepSeek-V3-2 & 18 (40.9\%) & \$0.21 & 60\% & 69\% & 54\% & 62\% \\
 \cline{2-10}
 &\multirow{4}{*}{\makecell[c]{Class\\Only}\qquad\makecell[c]{Single\\\\All}} &  \multirow{2}{*}{4} & Claude-4.5-Sonnet & 4 (100.0\%) & \$1.97  & 87\% & 100\% & 83\% & 75\%    \\
 && & DeepSeek-V3-2 & 1 (25.0\%) & \$0.24 & 37\% & 50.0\% & 33\% & 37\%  \\
 \cline{3-10}
&&\multirow{2}{*}{9} & Claude-4.5-Sonnet & 6 (66.6\%) & \$2.14 & 70\% & 75\% & 73\% & 63\%   \\
 && & DeepSeek-V3-2 & 2 (22.2\%) & \$0.18 & 46\% & 57\% & 48\% & 45\%  \\
 \cline{2-10}
&\multirow{2}{*}{\makecell{Mixed}} & \multirow{2}{*}{72} & Claude-4.5-Sonnet & 10 (13.8\%) & \$1.95 & 55\% & 78\% & 31\% & 65\%  \\
& &&  DeepSeek-V3-2 & 8 (11.1\%) & \$0.29 & 46\% & 65\% & 26\% & 53\% \\
 \cline{2-10}
&\multirow{2}{*}{\makecell{Total}} & \multirow{2}{*}{128} & Claude-4.5-Sonnet & 43 (33.5\%) & \$1.86 & 62\% & 82\% & 44\% & 69\%  \\
& & & DeepSeek-V3-2 & 28 (21.8\%) & \$0.25 & 50\% & 67\% & 37\% & 57\% \\
\midrule\midrule
\multirow{14}{*}{\makecell{\rotatebox{90}{SWE-PolyBench (Java)}}}
& \multirow{2}{*}{None} & \multirow{2}{*}{0} & Claude-4.5-Sonnet & -- & -- &  -- & -- & -- & -- \\
&& & DeepSeek-V3-2 & -- & -- & -- & -- & -- & --  \\
 \cline{2-10}
 % &\multirow{4}{*}{\begin{tabular}{c|c}
 %      \multirow{4}{*}{\makecell[c]{Function Only}}&\multirow{2}{*}{\makecell[c]{Single}}  \\
 %      \cline{2-2}
 %      &\multirow{2}{*}{Total}  \\
 %      \end{tabular}}&  \multirow{2}{*}{??} & Claude-4.5-Sonnet & &  &  &  &  &    \\
 &\multirow{4}{*}{\makecell[c]{Function\\Only}\quad\makecell[c]{Single\\\\All}} &  \multirow{2}{*}{31} & Claude-4.5-Opus & 18 (58.0\%) & \$1.53 & 84\% & 85\% & 74\% & 78\%    \\
 && & DeepSeek-V3-2 & 10 (32.2\%) & \$0.23 & 49\% & 50\% & 39\% & 40\%  \\
 \cline{3-10}
&&\multirow{2}{*}{46} & Claude-4.5-Opus & 25 (54.3\%) & \$1.61 & 77\% & 83\% & 69\% & 78\%    \\
 && & DeepSeek-V3-2 & 16 (34.7\%) & \$0.23 & 54\% & 62\% & 46\% & 53\%  \\
 \cline{2-10}
 &\multirow{4}{*}{\makecell[c]{Class\\Only}\qquad\makecell[c]{Single\\\\All}} &  \multirow{2}{*}{5} & Claude-4.5-Opus & 2 (40.0\%) &  \$3.35 &  90\% & 81\% & 40\% & 28\%    \\
 && & DeepSeek-V3-2 & 1 (20.0\%) & \$0.38 & 46\% & 50\% & 16\% & 30\%  \\
 \cline{3-10}
&&\multirow{2}{*}{10} & Claude-4.5-Opus & 4 (40.0\%) & \$3.88 & 74\% & 90\% & 44\% & 58\%    \\
 && & DeepSeek-V3-2 & 1 (10.0\%) & \$0.48 & 39\% & 55\% & 22\% & 43\%  \\
 \cline{2-10}
&\multirow{2}{*}{\makecell{Mixed}} & \multirow{2}{*}{109} & Claude-4.5-Opus & 26 (23.8\%) & \$2.82 & 53\% & 83\% & 28\% & 75\%  \\
& &&  DeepSeek-V3-2 & 13 (11.9\%) & \$0.29 & 34\% & 50\% & 19\% & 44\%  \\
 \cline{2-10}
&\multirow{2}{*}{\makecell{Total}} & \multirow{2}{*}{165} & Claude-4.5-Opus & 55 (33.3\%) & \$2.55 & 61\% & 84\% & 40\% & 75\%  \\
& & & DeepSeek-V3-2 & 30 (18.1\%) & \$0.28 & 40\% & 54\% & 26\% & 46\%  \\
\bottomrule
\end{tabular}
}
\end{table}

Some benchmarks come with instance labels that partition their set of
instances into different subsets.
Comparing metrics on these subsets yields further insight into agent
performance, while also helping us understand the benchmarks better.

Multi-SWE-bench~\cite{zan_et_al_2025} comes with hand-labeled
complexity levels~(Easy, Medium, or Hard).
\Cref{tab:complexity_hand_annotated} shows that the metrics follow the
expected trend: as issues get harder, \mbox{\% Resolved} decreases,
cost increases, and retrieval metrics decrease.
Of course, when metrics are computed on a smaller subset of
instances, they get more noisy.
SWE-PolyBench~\cite{rashid_et_al_2025} comes with hand-labeled
Verified tags for some instances whose issue
description~$d_\textrm{issue}$ is clear and matches its
tests~$t_\textrm{golden}$ well.
SWE-PolyBench~\cite{rashid_et_al_2025} indicates that on average, 
Verified instances are easier than non-Verified ones.
This implies that the numbers are not directly comparable across the
SWE-PolyBench leaderboards for Verified vs.\ Full.

Besides hand-labeled complexity levels, SWE-PolyBench also comes with
localization-derived labels that are indicative of complexity.
We wrote our own code for deriving these labels,
then ran it on both benchmarks.
\Cref{tab:complexity_loc_derived} shows the results.
Here, None means all changes are in new files or non-code files;
Function Only means the change is entirely contained in
functions~(which may or may not be part of the class);
Class Only means the change is entirely in a class~(e.g., changing
attributes, but not contained within a function of a class);
and Mixed counts any remaining changes across functions and classes or
top-level code in a file.
Rows marked Single only count instances where a single node was
modified, whereas rows marked All count instances where one or more
nodes of the given kind were modified.
Some of the subsets are small, leading to noisy results, but from the
more significant subsets we can observe trends.
In general, going from functions to classes to mixed, resolution rate
decreases, cost increases, and retrieval metrics decrease.

While Multi-SWE-bench~\cite{zan_et_al_2025} and
SWE-PolyBench~\cite{rashid_et_al_2025} are great contributions to the
research community, we also encountered some idiosyncrasies worth sharing.
Multi-SWE-bench instances come with \emph{identifier hints}, which are
newly-defined entity names, that could be used in addition to the issue
description~$d_\textrm{issue}$ to make the issue more specific.
Since we are unsure about whether or not other non-iSWE leaderboard
submissions make use of such identifier hints, we decided to hide them from
iSWE for our experiments, even if that reduces iSWE's apparent performance.
Some SWE-bench inspired benchmarks also have so-called \emph{hints-text}
that, despite the similar-sounding name, is unrelated to identifier hints.
Such hints-text comes from comments and discussions on the issue after
the original issue statement; using this is widely frowned upon, so
we hide it from iSWE for our experiments.
Multi-SWE-bench has \emph{duplicate issues} with identical issue
descriptions~$d_\textrm{issue}$ and golden code patches~$c_\textrm{new}$,
mostly due to back-porting fixes to different branches.
We treat these like any other instances for consistency with how
everyone else uses the benchmark.
Furthermore, Multi-SWE-bench also has some instances where
the \emph{golden code patch fails the tests}; again, we treat these
like any other instances, although not surprisingly, the code patch
generated by iSWE also fails the tests, so these remain unresolved.
Finally, since iSWE does not have a bash tool, it cannot cheat by
using commands such as \mbox{`\texttt{\small git log --all}'}
or \mbox{`\texttt{\small git show }\textit{<commit\_id>}'} that have been reported
from non-iSWE solutions on some SWE-bench derived leaderboards.

\begin{table}[t]
\centering
\caption{Performance by number of identifier hints about new entities provided in Multi-SWE-Bench (Java), using Claude-4.5-Sonnet.}
\label{tab:identifier_hints}
{\small
\begin{tabular}{c|r|r|rr|rr}
\toprule
\multirow{2}{*}{\makecell{\# Hints \\Provided}} & \multirow{2}{*}{\makecell{\#\\Instances}} & \multirow{2}{*}{\% Resolved} & \multicolumn{2}{c|}{File Retrieval} & \multicolumn{2}{c}{Node Retrieval} \\
\cmidrule(lr){4-5} \cmidrule(lr){6-7}
 & & & Recall & Precision & Recall & Precision \\
\midrule
0     & 79 & 44\% &  68.33\% &  84.99\% & 52.03\% &  73.89\% \\
1     & 23 & 30\% &  59.66\% &  86.96\% & 35.98\% &  59.50\% \\
2     &  4 & 25\% &  50.00\% & 100.00\% & 43.15\% &  95.00\% \\
3     &  2 & 0\%  &  50.00\% &  62.50\% & 50.00\% &  66.67\% \\
4     &  8 & 0\%  &  62.35\% &  75.00\% & 18.31\% &  68.84\% \\
5     &  2 & 0\%  & 100.00\% &  50.00\% & 66.67\% &  20.56\% \\
$\geq6$ & 10 & 0\%  &  15.18\% &  57.00\% & 14.12\% &  45.85\% \\
\bottomrule
\end{tabular}
}
\end{table}

To explore one of the above-listed idiosyncrasies more,
\Cref{tab:identifier_hints} shows iSWE's performance on instances
with different numbers of \emph{identifier hints}.
Since we are hiding identifier hints from iSWE, we expected iSWE to
perform worse on instances where the benchmark creators provided such
hints.
The results show that this expectation holds: iSWE's resolution rates
are highest for instances with zero hints, and higher for instances
with few hints than for instances with many.
The information-retrieval metrics are noisy, perhaps due to the small
subsets of instances.

\section{Related Work}\label{sec:related}

The authors of Multi-SWE-bench~\cite{zan_et_al_2025} and
SWE-PolyBench~\cite{rashid_et_al_2025} created leaderboards and
populated them using multi-lingual issue resolution systems they
adapted from prior work.
Since those adapted systems can handle Java code,
that makes them related work for our paper.
Specifically, the Multi-SWE-bench leaderboard has entries for
Magentless~(based on Agentless~\cite{xia_et_al_2025}
  but skipping test-based validation);
MSWE-agent~(based on SWE-Agent~\cite{yang_et_al_2024});
and MopenHands~(based on OpenHands~\cite{wang_et_al_2025}).
Similarly, the SWE-PolyBench leaderboard has entries for
Agentless-PB~(based on Agentless~\cite{xia_et_al_2025}
  but using only some parts of test-based validation);
SWE-agent-PB~(based on SWE-Agent~\cite{yang_et_al_2024});
and Aider-PB~(based on Aider~\cite{aider}
  but disabling interactive parts and test-based validation).
All of these systems have only limited language-specific components,
such as using Tree-Sitter~\cite{clem_thomson_2022} to create a
repository map in Aider and Agentless.
In contrast, iSWE puts a special emphasis on Java, with more advanced
Java-specific tools.

At the time of this writing, InfCode using GPT-5.2 is at the top of
the Multi-SWE-bench~(Java) leaderboard~\cite{li_et_al_2025}.
It works by alternating between two ReAct agent, a test generator and
a patch generator, causing higher cost;
in contrast, iSWE does not yet use test feedback.
The leaderboard for SWE-PolyBench Verified also has highly ranked
third-party submissions, including Java results beyond the entries
from the benchmark authors.
Prometheus~\cite{chen_et_al_2025} is an issue resolution system with five
sub-agents, all revolving around a common knowledge graph constructed
using Tree-Sitter.
In contrast, iSWE uses a more direct tool-based approach with no
knowledge graph and more emphasis on Java.
Furthermore, our paper reports more detailed results, such as
localizer metrics not present in the Prometheus paper.
The other third-party leaderboard entries on SWE-PolyBench with Java
results are Atlassian Rovo Dev~\cite{rovodev} and Amazon Q Developer
Agent~\cite{amazon_q}.
Unfortunately, their inner workings are unknown; while there is a
paper about a predecessor system of Rovo
Dev~\cite{takerngsaksiri_et_al_2025}, it does not talk about its
tools, and even if it did, Rovo Dev may work differently.
We can thus not compare their approach, only their results
(see \Cref{sec:evaluation}).

While the above discussion focuses on Java issue resolution, so far,
more of the action has been on Python issue
resolution~\cite{aider,chen_et_al_2024,chen_et_al_2025,gao_et_al_2025,tao_et_al_2024,wadhwa_et_al_2024,wang_et_al_2025,xia_et_al_2025,yang_et_al_2024,zhang_et_al_2024}.
One dimension for the design of issue resolution systems is whether to
use zero, one, or multiple agents.
Agentless~\cite{xia_et_al_2025} uses zero ReAct agents, and instead,
has a fixed workflow, both at the top-level and within its sub-modules
for localization and code editing.
Single-agent systems, such as SWE-Agent~\cite{yang_et_al_2024},
Aider~\cite{aider}, or OpenHands~\cite{wang_et_al_2025}, use a unified
ReAct loop to handle both localization and code editing.
Multi-agent issue resolution systems rely upon anywhere from
two~(AutoCodeRover~\cite{zhang_et_al_2024}, HULA~\cite{takerngsaksiri_et_al_2025})
to four~(CodeR~\cite{chen_et_al_2024}, Magis~\cite{tao_et_al_2024})
or even five~(MASAI~\cite{wadhwa_et_al_2024}, Prometheus~\cite{chen_et_al_2025})
sub-agents.
As discussed in \Cref{sec:approach}, iSWE has two sub-agents
with a hand-off based on a sanitized and well-typed set of locations.

More important than the number of sub-agents may be the tools being
called by these agents.
Some systems leverage basic file system tools, specifically bash shell
commands for code search and a single custom tool to view and edit files.
Since the release of Claude~3.5 Sonnet, several systems have converged
to a \emph{str\_replace\_editor} tool to view files and make changes
(create and edit) to existing
files~\cite{gao_et_al_2025,wadhwa_et_al_2024,wang_et_al_2025,xia_et_al_2025,yang_et_al_2024}.
This is true even for some systems that previously used their own
different tools, since Claude is a strong code model specifically
trained with these tools in mind.
Another approach is to introduce more sophisticated, domain-specific
tools.
For example, AutoCodeRover~\cite{zhang_et_al_2024} employs AST-based
tools that enable searching for specific code entities~(classes,
methods) within other entities, and
Prometheus~\cite{chen_et_al_2025} builds a code knowledge graph for
the code repository, allowing the agent to search for classes and
functions.
With iSWE, we double down on more sophisticated tools specialized for Java.
At the same time, iSWE differs from most prior issue resolution
systems by being careful to use mostly read-only tools.

One topic this paper did not cover is testing and inference scaling.
Several issue-resolution systems run existing regression tests or
newly-generated reproduction tests, either to refine~$c_\textrm{new}$
or to select among multiple candidates
for~$c_\textrm{new}$~\cite{ehrlich_et_al_2025,gao_et_al_2025,jain_et_al_2025,xia_et_al_2025,zhang_et_al_2024}.
In fact, this practice can be so useful that our team created a
dedicated benchmark for reproduction test generation, TDD-Bench
Verified~\cite{ahmed_et_al_2024}.
But while we have ongoing work on both testing
(e.g.~\cite{ahmed_et_al_2026}) and inference scaling, that work is
beyond the scope of this paper, which focuses on the core agent.

% The fourth category introduces more sophisticated, domain-specific tools. AutoCodeRover \cite{autocoderoverpaper} employs AST-based tools that enable searching for specific code elements (classes, methods) within other code elements. TRAE \cite{gao_et_al_2025} builds code knowledge graph for the code repository, allowing the agent to search for classes and functions. Code Graph Model (CGM) \cite{tao2025code} also builds a code graph but uses Graph RAG to identify the most relevant files for modification. MarsCode Agent \cite{liu2024marscode} combines code knowledge graphs and language server protocols to provide agents with capabilities for code entity retrieval, relationship retrieval, and definition-and-reference navigation, enabling agents to browse and analyze code similarly to human developers.

% A third important dimension is the training approach, with recent work exploring fine-tuned models using supervised and reinforcement learning techniques. 
% SWE Fixer \cite{xie2025swe} trains models specifically for localization and coding tasks, achieving competitive results using only 2 inferences per issue. MCTS-Refine-7B \cite{mctsrefine7B} incorporates Monte Carlo Tree Search with self-improvement mechanisms and multi-agent debate. SWE-RL \cite{wei2025swe} and DeepSWE \cite{deepswe} achieve strong performance with scaling through RL training. 

Recently, there has been significant effort to identify areas where issue resolution systems excel and where they perform poorly. Analysis of agent performance on multi-file issues \cite{ganhotra2025multifile} and single-file saturation studies \cite{ganhotra2025saturation} reveal that state-of-the-art systems struggle significantly with multi-file issues, despite achieving high performance on single-file problems. Further research, such as TRAIL~(Trace Reasoning and Agentic Issue Localization)~\cite{deshpande2025trail} and MAST~(Multi-Agent System Failure Taxonomy)~\cite{cemri2025multi}, explores the failure modes of agents by analyzing their trajectories, identifying various reasons for failure including incorrect fault localization, inability to understand complex code dependencies, and challenges in maintaining context across multiple files. 
These studies highlight the gap between current agent capabilities and the requirements of real-world software engineering tasks, where issues frequently span multiple files and require deep understanding of system architecture.

\section{Conclusion}\label{sec:conclusion}

This paper introduces iSWE, a two-agent system for resolving issues on
code repositories with a particular emphasis on Java.
The first sub-agent uses read-only tools based on static analysis to
find a set of code locations to be edited.
The second sub-agent uses an edit tool, which also performs most of
its work in a read-only manner, with any side-effects
isolated in a container.
The tools for both agents leverage Java knowledge for better performance.
Our results show that iSWE reaches state-of-the-art issue resolution
rates on the Java splits of two benchmarks, SWE-PolyBench and
Multi-SWE-Bench.
It achieves these results without any model fine-tuning or dynamic
test execution feedback.
We expect that incorporating those in future work will further
increase resolution rates.

%% \begin{acknowledgements}
%%   % content in acknowledgements will be automatically hidden during submission
%% \end{acknowledgements}

\bibliography{references}

%% \newpage
%% \appendix

%% % \supplemental material can be placed here; this material will be hidden if the
%% % [hidesupplement] option is provided

%% \section{Example Supplemental Section}
%% % This example section may be removed.

\end{document}